\documentclass[letterpaper,journal]{IEEEtran}
\PassOptionsToPackage{expansion=false}{microtype}
\usepackage{amsmath}
\usepackage{amssymb}
\usepackage{bm}
\usepackage{booktabs}
\usepackage{tabularx}
\usepackage{array}
\usepackage{caption}
\usepackage{enumitem}
\usepackage{graphicx}
\usepackage{placeins}
\usepackage[most]{tcolorbox}
\usepackage{cite}
\usepackage{balance}
\usepackage[hidelinks]{hyperref}

\graphicspath{{Figures/}}
\AtBeginDocument{%
  \captionsetup[figure]{font=small}%
}

\newcommand{\edgeAcc}{74.25\%}
\newcommand{\cloudAcc}{90.81\%}
\newcommand{\eceCal}{2.296\%}
\newcommand{\eceUncal}{14.757\%}
\newcommand{\opThreshold}{0.72}
\newcommand{\temperature}{2.13}
\newcommand{\cleanOffload}{45.99\%}
\newcommand{\cleanCombinedAcc}{89.83\%}
\newcommand{\cloudCorrection}{37.29\%}
\newcommand{\phaseTransition}{0.50}
\newcommand{\latencyClean}{13.91\,ms}
\newcommand{\latencyAttacked}{28.69\,ms}
\newcommand{\latencyAmp}{2.062$\times$}

\newcommand{\cloudWorkAmp}{2.174$\times$}
\newcommand{\benignPninetynineAmp}{10.46$\times$}
\newcommand{\cleanBenignPninetynine}{166\,ms}
\newcommand{\attackBenignPninetynine}{1737\,ms}
\newcommand{\beMalSybil}{69\%}
\newcommand{\tbConcPninetynine}{252\,ms}
\newcommand{\gtbMalSybil}{59\%}
\newcommand{\gtbUnauthLossSybil}{52.1\%}
\newcommand{\gtbAuthLossMax}{1.29\%}
\newcommand{\beUnauthLossSybil}{2.8\%}
\newcommand{\benignAccNoDef}{89.7\%}
\newcommand{\benignAccGtbSybil}{85.5\%}
\newcommand{\benignAccDropSybil}{4.2}
\newcommand{\tbSybilPninetynine}{1730\,ms}
\newcommand{\cleanAuthPninetynine}{91\,ms}
\newcommand{\cleanAuthLossBE}{1.25\%}

\newcommand{\cfRhoClean}{0.63}

\newcommand{\cfPnnClean}{165}
\newcommand{\cfPnnPert}{1738}

\newcommand{\cfSeeds}{12}
\newcommand{\epsN}{1000}

\newcommand{\epsQb}{0.463}
\newcommand{\epsMinSat}{0.5}

\newcommand{\convOneStepTwo}{0.828}
\newcommand{\convOneStepFour}{0.850}
\newcommand{\convFiveStep}{0.998}
\newcommand{\pdN}{1000}

\newcommand{\pdNumDefenses}{12}

\newcommand{\aaMalShareMax}{59\%}
\newcommand{\aaUnauthRejMax}{52\%}

\newcommand{\aaSeeds}{12}

\newcommand{\dmAggNoneHundred}{94.9\%}

\newcommand{\dmAggNoneFiveHundred}{70.8\%}

\newcommand{\dmAllAccNone}{89.7\%}

\newcommand{\dmAggBEHundred}{46.8\%}

\newcommand{\dmAggBEFiveHundred}{25.0\%}

\newcommand{\dmAllAccBE}{89.5\%}

\newcommand{\dmAggGtbHundred}{0.5\%}

\newcommand{\dmAggGtbFiveHundred}{0.0\%}

\newcommand{\dmAllAccGtb}{85.5\%}
\newcommand{\dmAggNoneZeroHundred}{7.8\%}

\newcommand{\revCPUSeeds}{8}
\newcommand{\revCPUHorizon}{800}
\newcommand{\revCPUWarmup}{100}

\begin{document}
\title{GateDrain: Availability Attacks and Admission-Side Defense for Confidence-Gated Edge--Cloud Inference}
\author{Zonghua Gu, Julian Singh-Smith, Junlin Liao, and Di Liu
\thanks{Z. Gu, J. Singh-Smith, and J. Liao are with Hofstra University, Hempstead, NY, USA. E-mail: zonghua.gu@hofstra.edu.}
\thanks{D. Liu is with the Norwegian University of Science and Technology, Norway. }
}
\maketitle

\begin{abstract}
Confidence-gated edge--cloud inference accepts confident local predictions and
offloads uncertain inputs to a stronger cloud model. We show that this routing
decision creates an availability attack surface. We call this attack
\emph{GateDrain}: bounded input perturbations lower calibrated confidence and
redirect requests that would otherwise be answered locally into a shared
cloud queue, without increasing the application request rate. Because escalated requests share a
cloud service, an increase in per-request cloud demand can move a
near-capacity deployment across a queueing knee, causing disproportionate
tail-latency degradation for benign users. We evaluate white-box, transfer,
decision-only, universal, and multi-gate attacks on the public EdgeBoost
artifact. A fixed-application-volume comparison isolates the effect of confidence
manipulation from added client traffic, while perturbation-budget and
arrival-process sweeps show that the queueing transition persists across
several workload models but its amplification depends on the operating point. Adaptive attacks also defeat
the evaluated training-free preprocessing defenses. To contain the resulting
cloud demand, we evaluate Bounded Escalation, which combines per-source
admission budgets, protected capacity, and non-preemptive trusted-class
priority; an optional global bucket adds an identity-independent bound on
untrusted admissions. The evaluation makes the resulting policy trade-off
explicit: authenticated clients receive latency isolation, whereas tighter
aggregate containment can reject legitimate unauthenticated offloads and
reduce overall expected accuracy through edge fallback. 
\end{abstract}

\begin{IEEEkeywords}
edge--cloud inference,
selective offloading,
model cascades,
adversarial availability,
edge AI security
\end{IEEEkeywords}

\section{Introduction}
\label{sec:intro}

Resource-constrained Internet of Things (IoT) devices increasingly rely on
edge--cloud offloading to balance execution latency, communication cost,
energy consumption, reliability, and quality of service (QoS). Recent works
use workload and resource-state prediction to coordinate cloud and edge
resources in industrial IoT~\cite{sun2021cloud}, whereas TORE-NET combines
task replication and offloading to satisfy reliability and energy
constraints~\cite{rasouli2026tore}. These studies treat offloading as a
performance and resource-management decision.

Confidence-gated edge--cloud inference makes the offloading decision from
an input-derived model-confidence signal. It accepts confident local
predictions and escalates uncertain inputs to a stronger cloud model.
\emph{EdgeBoost}\cite{edgeboost2025} calibrates a small edge model (e.g., MobileNetV3-Small)
\cite{mobilenetv3} with temperature scaling and pairs it with a
large cloud model (e.g., EfficientNetV2-L)~\cite{efficientnetv2}. The calibrated confidence
signal is therefore a differentiable routing control point: manipulating it
changes where inference is performed even when cloud inference can recover a
useful prediction.

When an offloading decision depends on such a confidence signal, it also
becomes a security-sensitive control surface. We call this confidence-gate
escalation attack \emph{GateDrain}: it redirects otherwise local inference
to a shared backend and increases cloud contention without increasing the
application request rate. White-box, transfer, decision-only, and universal
settings (\S\ref{sec:attacks}) instantiate GateDrain under progressively
weaker adversary knowledge; its primary setting pushes the
calibrated top-1/top-2 margin below the offload threshold. In the primary
experiment, offload rises from $46.3\%$ to $99.9\%$ ($2.16\times$
cloud-demand amplification). In a separate
evaluation using an independently trained cloud checkpoint, the attack
reaches $99.9\%$ offload and reduces cloud accuracy on the same
$999$-image population from $905/999$ ($90.6\%$) on clean versions to
$823/999$ ($82.4\%$) on perturbed versions (\S\ref{sec:eval-setup}).
Near cloud
capacity, the resulting demand increase causes
disproportionate queueing delay for benign requests sharing the service. The
observed \benignPninetynineAmp{} admitted-benign p99 amplification is specific
to the Poisson-arrival baseline of our $50$\,req/s, serial batch-one,
$K{=}64$ case study, not a universal multiplier.

A confidence gate controls both a prediction-routing decision and admission to
a shared resource. Improving prediction robustness and containing admitted
resource demand therefore address different questions. We propose
\emph{Bounded Escalation}, an admission-side containment policy that limits
cloud work after routing without requiring the perturbation to be detected.
It combines per-source token budgets, protected capacity, and non-preemptive
trusted-class priority. An optional global bucket adds an
identity-independent bound on aggregate untrusted admissions. This separation
is important under Sybil traffic: per-source budgets bound each identity but
do not bound the aggregate work of many identities. The optional global bucket
provides aggregate containment at a substantial utility cost to legitimate
unauthenticated clients.

\begin{figure}[t]
\centering
\includegraphics[width=\columnwidth]{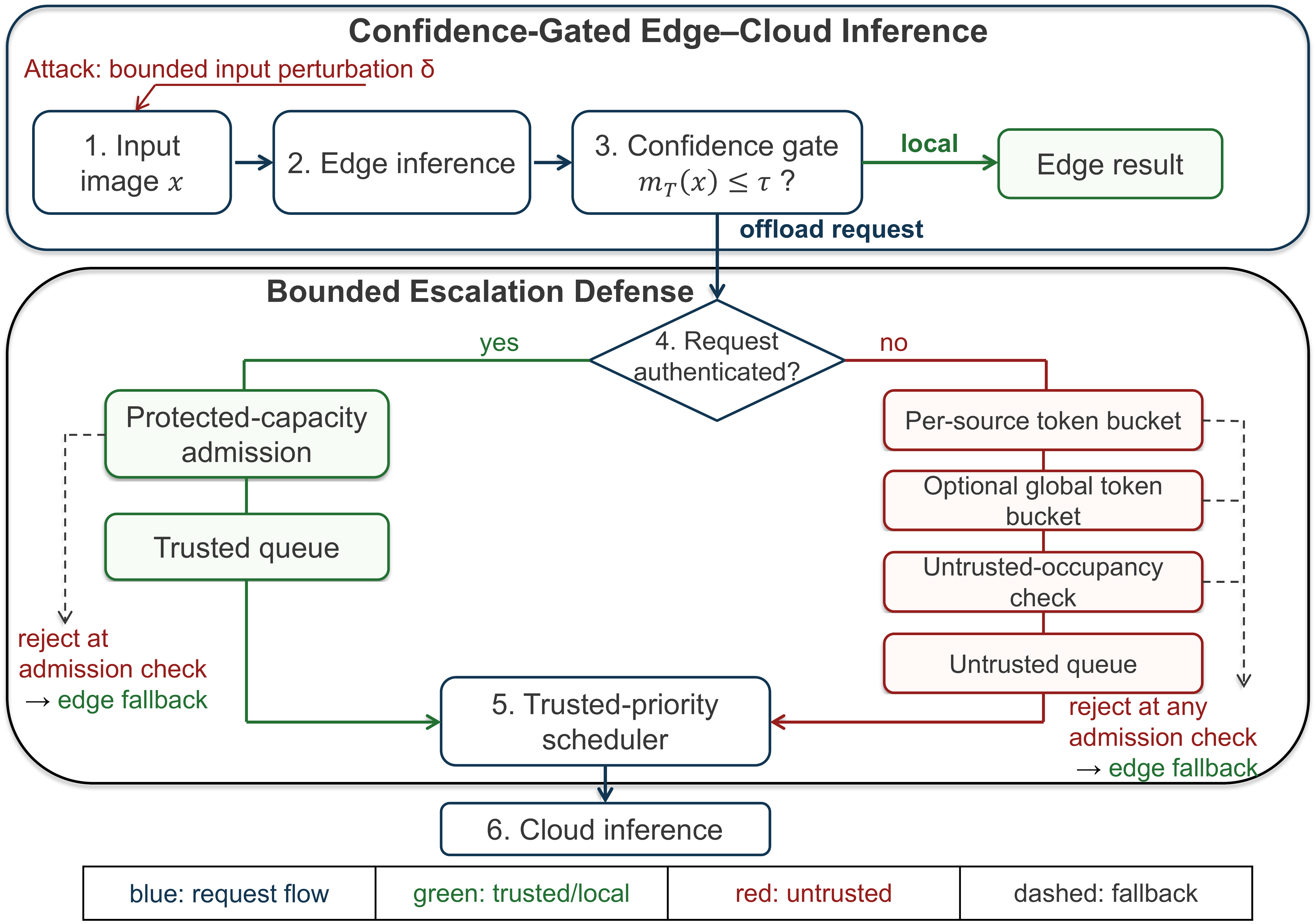}
\caption{Top: A bounded perturbation $\delta$ modifies the input image and
manipulates the confidence gate, generating unnecessary offload requests.
Bottom: Every offload request is intercepted by Bounded Escalation before
reaching the cloud. Authenticated requests use total-capacity admission and the
trusted queue, while unauthenticated requests must pass admission checks before
entering the untrusted queue. A request rejected at any admission stage returns
the edge prediction already computed before the offload decision.}
\label{fig:overview}
\end{figure}

Although EdgeBoost~\cite{edgeboost2025} provides our concrete case study, \emph{the
threat model and defense address a broader architectural pattern}: any
edge--cloud system in which an input-dependent confidence signal controls
admission to a more expensive shared backend. Figure~\ref{fig:overview}
summarizes the escalation attack and admission defense.

Our contributions are:
\begin{enumerate}[leftmargin=*,nosep]
    \item We formulate calibration-aware confidence-gate attacks under
    white-box, transfer, routing-decision-only, and universal settings, and
    evaluate three confidence-gate signals.

    \item We connect attack-induced offloading to shared-cloud queueing using
    measured A100 service-time samples. A fixed-application-volume comparison
    isolates the effect of confidence manipulation from added client traffic,
    while perturbation-budget, service-model, arrival-process, and
    operating-point sensitivity analyses characterize the conditions under
    which tail amplification occurs.

    \item We evaluate twelve training-free input-preprocessing baselines under
    oblivious and adaptive attacks on the complete artifact split, including
    BPDA and EOT attack adaptations.

    \item We evaluate Bounded Escalation under concentrated, Sybil,
    rate-shaped, and duty-cycled floods, reporting authenticated isolation,
    aggregate containment, legitimate unauthenticated rejection, and
    edge-fallback accuracy costs.
\end{enumerate}

The remainder of this paper is organized as follows.
Section~\ref{sec:background} reviews related latency and offloading attacks.
Section~\ref{sec:threat} defines the confidence gate, queueing model, and
attacker capabilities. Section~\ref{sec:attacks} presents the escalation
attacks, and Section~\ref{sec:queue} analyzes attack-induced queue
amplification. Section~\ref{sec:defense} presents Bounded Escalation.
Section~\ref{sec:eval} evaluates attacks, defenses, arrival processes,
operating points, and admission trade-offs. Finally,
Section~\ref{sec:conclusion} concludes.

\section{Background and Related Work}
\label{sec:background}

\subsection{Dynamic-Inference Slowdown Attacks}

Latency attacks target availability by increasing inference-time
computation, energy, or response time without necessarily changing the
predicted label~\cite{brachemi2026energy}; a cross-domain survey
organizes them by exploited bottleneck~\cite{gu2026latency}.
Representative examples suppress early exits~\cite{deepsloth2021},
reactivate skipped blocks~\cite{ilfo2020}, extend slowdown attacks to
black-box settings~\cite{splat2026}, reduce activation
sparsity~\cite{spongeexamples2021}, or inflate tracker
non-maximum-suppression work~\cite{overload2024}, with demonstrated
downstream safety consequences for real-time perception~\cite{slowtrack2024}.
These attacks amplify computation within one inference path; ours
instead targets the edge--cloud resource-allocation boundary and
amplifies cloud admissions and shared-queue latency.

Existing defenses attempt to keep adversarial inputs off expensive
execution paths: hardware-adaptive adversarial training restores
detector throughput under latency attacks~\cite{wang2025cant}, and DefQ
applies defensive quantization before a multi-exit model~\cite{qiu2023defq},
conjecturing that gradient-obfuscation countermeasures survive adaptive
attacks; our Section~\ref{sec:preddef} results bear on that conjecture in
the confidence-gate setting. We evaluate such non-differentiable defenses
adaptively using expectation over transformation (EOT, averaging the
attack loss over transformation samples) and the backward-pass
differentiable approximation (BPDA, a differentiable surrogate gradient
in the backward pass). Our defense is complementary: it neither detects
nor removes the perturbation, but bounds the cloud work admitted after
the confidence gate.

\subsection{Confidence-Guided Adaptive Inference}
\label{sec:related-adaptive}

Input-dependent resource allocation predates modern edge--cloud
inference: the Viola--Jones cascade spends computation only on image
regions that survive cheap early tests~\cite{viola2001rapid}, but its
decisions concern regions within one detector rather than calibrated
confidence routing between an edge device and a shared cloud service.

Several lines of work make computation input-dependent without
instantiating the attack surface studied here. Adaptive policies select
among or terminate pretrained networks using confidence features,
including edge--cloud deployments, but do not consider adversarial
manipulation of the selection policy or queue
contention~\cite{bolukbasi2017adaptive}. A recent line of confidence-based
model cascades for large language models formalizes the same
try-cheap-then-escalate pattern to optimize cost and accuracy within a
single serving stack~\cite{chen2024frugalgpt}; our setting instead spans a
trust boundary between an edge device and a shared cloud service, and our
concern is adversarial manipulation of the escalation decision rather than
its cost-optimality. Selective classification
abstains on low-confidence inputs~\cite{geifman2017selective}, which
allocates no additional computation unless a deployment attaches costly
downstream processing to abstention. Early-exit
networks~\cite{teerapittayanon2016branchynet,huang2018multiscale} place
the adaptive decision inside one model, so their principal cost is local
execution rather than cloud admission. Neurosurgeon partitions a DNN by
layer profiles and network conditions rather than per-prediction
confidence~\cite{kang2017neurosurgeon}.

Collectively, these studies show that an input-dependent decision can control
whether additional computation is performed, but the security consequence
depends on what lies beyond the decision. Additional local layers increase
per-request computation; abstention may allocate no additional resource; and
cloud escalation adds communication, admission, and shared queueing. Our work
does not claim that every adaptive-inference system above is directly
vulnerable. It focuses on the specific case in which bounded image
perturbations can manipulate a calibrated model-confidence gate, convert
ordinary application requests into cloud work, and amplify latency through a
shared queue. Bounded Escalation consequently limits admissions per identity
and across the untrusted tier, rather than modifying the underlying adaptive
model architecture.

\subsection{Offloading Attacks and the Gate Distinction}
Offloading attacks manipulate either the routing policy or the
infrastructure capacity behind it. The Stealthy Interference
Attack~\cite{zhang2020fooling} uses low-power wireless interference to
alter the SINR states observed by a DRL offloading policy, redirecting
tasks to an attacker-selected server; it targets server selection and
data confidentiality rather than server overload.

Capacity-removal attacks~\cite{tavori2026worst} selectively disable
edge servers so that a reactive operator must redistribute load to the
cloud; convexity analysis shows worst-case attacks concentrate budget on
critical sites, changing offloading indirectly through capacity loss.

Prior work manipulates intra-model execution, observed wireless state,
or available server capacity; our attack instead manipulates the
calibrated image-confidence gate that controls cloud admission, leaving
the infrastructure and application request rate unchanged, so the
resulting increase in cloud admissions amplifies shared-queue latency
rather than per-input local computation.

\section{System and Threat Model}
\label{sec:threat}

\subsection{Gate Model}
Temperature scaling~\cite{guo2015calibration} is a standard post-hoc
calibration method: it rescales the model's logits $z$ by a single scalar,
$p(x)=\mathrm{softmax}(z/T)$, which reduces overconfidence without altering
the argmax prediction or accuracy. We do not train or fit this calibration
ourselves; EdgeBoost ships a model already calibrated with an undisclosed
$T$, and we recover $T{\approx}\temperature$ by solving for the value that
reproduces the shipped calibrated probabilities from the shipped
uncalibrated logits. Section~\ref{sec:eval} confirms the calibration is
effective: ECE drops from \eceUncal{} to \eceCal{} with top-1 labels
unchanged. Let $p(x)$ denote these calibrated probabilities, and let
$p_{(1)}(x)$ and $p_{(2)}(x)$ be the largest two entries. The \emph{margin} and \emph{offload decision} are
\begin{equation}
  m_T(x) = p_{(1)}(x) - p_{(2)}(x),
  \label{eq:margin}
\end{equation}
\begin{equation}
  o(x) = \mathbb{1}\!\left[\, m_T(x) \le \tau \,\right],
  \label{eq:gate}
\end{equation}
where $o(x)=1$ denotes \emph{escalation}, i.e., offloading from device
to cloud. The attack differentiates through this deployed gate. We also
evaluate max-softmax and entropy gates.

\subsection{Queue Model}
Benign and adversarial requests arrive at rates $\lambda_b$ and $\lambda_a$
with offload probabilities $q_b=\mathbb{E}[o(x_b)]$ and
$q_a=\mathbb{E}[o(x_a{+}\delta)]$. Offered cloud load is
\begin{equation}
  \lambda_c = \lambda_b q_b + \lambda_a q_a.
  \label{eq:cloudrate}
\end{equation}
For service rate $\mu$, utilization is $\rho=\lambda_c/\mu$. We use
standard Kendall notation $A/B/c/K$, where $A$ and $B$ denote the arrival and
service-time distributions ($M$: memoryless/exponential, $D$:
deterministic, $G$: general/empirical), $c$ the server count, and $K$ the
system capacity. An infinite-buffer
M/M/1 reference gives $W_q=\rho/[\mu(1-\rho)]$ for $\rho<1$, illustrating the
sharp increase near capacity. Our results instead use a finite-capacity
measured-service queue: excess load causes rejection, not unbounded queue
growth. Offered load and admitted load therefore differ under saturation and
admission control.

\subsection{Attacker Capabilities and Security Goals}
The attacker perturbs image inputs within a \emph{bounded perturbation}, an $\ell_\infty$ budget in pixel space and controls
one or more unauthenticated identities. White-box attackers know the edge
model and calibrator; transfer attackers use a surrogate; decision-only
attackers observe the routing bit. We also evaluate an offline universal
perturbation. The objective is to increase cloud demand and benign tail
latency, not necessarily to change final labels.

The attacker cannot modify the models, queue, authentication service, or
hardware. Authentication determines protection eligibility, not maliciousness:
benign unauthenticated clients share the untrusted tier with attackers.
Compromised credentials and arbitrary authenticated floods are outside the
model. Bounded Escalation protects authenticated headroom and service order;
its optional global bucket bounds aggregate untrusted admissions, but neither
identifies attackers nor guarantees service to unauthenticated clients.

Bounded perturbations model an adversary constrained to modifying a
prescribed digital input while limiting visible deviation. Arbitrary input
selection represents a more permissive threat model and can induce offloading
without perturbation; we evaluate it separately in \S\ref{sec:selection}. Neither experiment establishes physical robustness through camera capture, compression, or resizing.

\section{Escalation Attacks}
\label{sec:attacks}

The primary attack minimizes a hinge loss against the calibrated margin:
\begin{equation}
  \mathcal{L}_{\text{esc}}(x') = \max\!\big(m_T(x')-\tau+\kappa,\;0\big).
  \label{eq:hinge}
\end{equation}
The slack $\kappa>0$ encourages crossing below $\tau$; the loss vanishes at
$m_T(x')\le\tau-\kappa$. Projected gradient descent
(PGD)~\cite{madry2018} differentiates through temperature scaling and
deployment preprocessing, projecting into the pixel-space $\ell_\infty$ ball.
Per-input white-box PGD is a strong reference attack, not a real-time edge-side
implementation; its optimization cost is excluded from online latency.
For max-softmax we minimize $\max_jp_j(x')$ below its threshold; for entropy
we maximize $H(p(x'))=-\sum_j p_j\log p_j$ (raw nats) above its threshold.

A \emph{universal} attacker learns one input-independent $\delta$ offline on a
disjoint training split over a range of thresholds, optionally penalizing label
change with a KL term, then replays it without online
optimization~\cite{deepsloth2021}. A \emph{transfer} attacker optimizes on a
surrogate MobileNet without victim gradients.

The \emph{routing-decision-only} attacker first queries the clean routing bit and stops
if the input already offloads. Otherwise it tests up to $30$ candidates
$\mathrm{clip}_{[0,1]}(x+r)$ with independent uniform sign vectors
$r\in\{-\epsilon,+\epsilon\}^{d}$, stopping at the first offload. This costs at
most $31$ oracle observations per image and uses no scores, gradients, or model
access. The paper's no-volume-increase claim concerns attack execution after
the perturbation is constructed: the decision-only search consumes up to $31$
routing observations per image, a separate query cost excluded from the queue
workload. Reliable routing observations are assumed, as in \S\ref{sec:background}.
This assumption is plausible in deployment: an offloaded request returns
after at least one cloud service time, whereas a local answer returns after
edge inference alone, and the measured gap (mean $27.4$\,ms cloud versus $1.30$\,ms edge) spans
more than an order of magnitude. The measured compute-time gap motivates
treating routing as an observable oracle; robustness to real network jitter
remains untested, and we do not build or measure this side channel.

Unless stated otherwise, PGD uses $\epsilon=8/255$, $40$ steps, step size
$1/255$, one random start, $\kappa=0.02$, and seed $42$.

\medskip
\noindent\textit{Intuition: escalate, then concentrate.}

\textbf{Attack mechanism.} The edge model (MobileNetV3-Small, $1.30$\,ms) answers
locally when it is confident and ships the input to the cloud
(EfficientNetV2-L, $27.4$\,ms) when it is not. ``Confident'' means the
calibrated margin between the top guess and the runner-up,
$m_T(x)=p^{(1)}-p^{(2)}$ \eqref{eq:margin}, stays above $\tau$; the gate
offloads when $m_T\!\le\!\tau$ \eqref{eq:gate}. The routing objective
does not explicitly require a label change; it only requires narrowing the
top-two margin. In practice, however, the resulting perturbation changes
many edge predictions, as quantified in \S\ref{sec:eval-setup}. The
attack of \eqref{eq:hinge} minimizes the hinge
$\mathcal{L}_{\mathrm{esc}}=\max(m_T-\tau+\kappa,0)$, which reaches zero once
$m_T\le\tau-\kappa$. We call this attack \emph{calibration-aware} because
it differentiates through the deployed temperature-scaling step that
produces $p(x)$ rather than treating the edge model's raw, uncalibrated
logits as the attack surface: the perturbation is optimized directly
against the margin $m_T(x)$ of Eq.~\eqref{eq:margin} that the gate actually
thresholds, not a proxy computed before calibration is applied. The perturbations are bounded in pixel space to limit visible distortion. No extra requests are sent: the attacker submits the
same application volume but increases the expected cloud work per request,
raising $q_a=\mathbb{E}[o(x_a+\delta)]$ toward one while the benign rate
stays at $q_b$ (\S\ref{sec:queue}).

\section{Cross-Layer Queue Amplification}
\label{sec:queue}

Writing $\varphi=\lambda_a/\Lambda$ for the adversarial fraction of the
offered application load, where $\Lambda=\lambda_a+\lambda_b$,
Eq.~\eqref{eq:cloudrate} becomes
\begin{equation}
  \lambda_c(\varphi)
  =
  \Lambda\big[(1-\varphi)q_b+\varphi q_a\big].
  \label{eq:cloudmix}
\end{equation}
When $q_a>q_b$, the offered cloud load increases with the adversarial
fraction. The corresponding service-capacity boundary is
\begin{equation}
  \varphi^\ast
  =
  \frac{\mu/\Lambda-q_b}{q_a-q_b},
  \label{eq:critical}
\end{equation}
provided $\varphi^\ast\in[0,1]$. Below this boundary, an increase in
offloading raises utilization and waiting time; near the boundary, a modest
increase in cloud demand can cause disproportionate tail latency. Under the
full-artifact clean baseline $q_b=0.4599$, saturating $q_a$ multiplies the
per-request expected cloud work by at most $1/q_b\approx2.174\times$. In
the finite-capacity system evaluated here, overload produces both delay and
rejection rather than unbounded queue growth.

The direct-inference experiment measures $q_a=0.999$, whereas the queue
sweeps use the saturated approximation $q_a=1$. At the primary operating
point this approximation changes offered cloud load by only
$0.025$\,req/s; we state it explicitly because the system is evaluated
near the service-capacity boundary. With $\Lambda=50$\,req/s,
$q_b=0.4599$, $q_a=1$, and $\mu\approx36.5$\,req/s,
Eq.~\eqref{eq:critical} gives $\varphi^\ast\approx\phaseTransition$.
Section~\ref{sec:eval} evaluates this cross-layer effect using measured
cloud service times and fixed application request volume.

\section{Bounded Escalation Defense}
\label{sec:defense}

Bounded Escalation composes admission and scheduling mechanisms at the
offload boundary (Fig.~\ref{fig:overview}). It does not require perturbation
detection. Each offload attempt passes a per-source token
bucket~\cite{tokenbucket}; authenticated requests may then use total capacity
$K$, while untrusted requests are admitted only if total occupancy is below
$K-K_p$. This prevents untrusted arrivals from consuming the protected
headroom, but does not keep those positions empty under authenticated load.
The implementation applies per-source budgeting to authenticated sources too,
so authentication does not imply zero rejection.

Admitted requests enter trusted and untrusted queues. Non-preemptive strict
priority selects a trusted request whenever one is waiting, with per-source
round-robin within each class. An untrusted job already in service is not
interrupted, and its residual service time still contributes to authenticated
latency.  Rejection at any admission stage triggers an \emph{edge fallback}:
the system returns the edge prediction that was already computed before
the confidence gate requested cloud offloading. This avoids additional
cloud work and queueing delay, but may sacrifice the accuracy improvement
that the cloud model would otherwise provide.

The optional global untrusted bucket charges each untrusted admission,
independently of identity. Its depth is $b_{\mathrm{gtb}}=20$ and nominal
refill rate $r_{\mathrm{gtb}}=1.3(25)(0.46)=14.95$/s at the matched-flood
operating point. Bounded Escalation contains offload-induced resource amplification
without attempting to classify requests as benign or malicious: admission
depends only on identity-scoped token budgets, tier membership, and queue
position, never on an estimate of whether traffic is anomalous. No
anomaly-detection or per-source rate-monitoring component is used.

For an observation window $\mathcal W$ and full initial buckets, a
per-source token bucket bounds any single identity's admitted offloads,
\begin{equation}
N^{\mathrm{adm}}_{u}(\mathcal{W}) \le B_u+R_u\mathcal{W},
\label{eq:bounds}
\end{equation}
but a Sybil attacker can spread its volume across $n_{\mathrm{id}}$
identities, and summing Eq.~\eqref{eq:bounds} over them gives only
\begin{equation}
\sum_{u=1}^{n_{\mathrm{id}}}N^{\mathrm{adm}}_{u}(\mathcal{W})
\le n_{\mathrm{id}}B_u+n_{\mathrm{id}}R_u\mathcal{W},
\label{eq:sybilbound}
\end{equation}
a bound that grows with the identity count. The optional global untrusted
bucket instead enforces the identity-independent bound
\begin{equation}
N^{\mathrm{adm}}_{\mathrm{untr}}(\mathcal{W}) \le
b_{\mathrm{gtb}}+r_{\mathrm{gtb}}\mathcal{W},
\label{eq:gtbbound}
\end{equation}
which applies only with the global bucket enabled. For the defense
analysis, we model all-benign accuracy from clean route-conditional
correctness: local answers use the clean edge prediction, admitted offloads
use the clean cloud prediction, and rejected offloads return the clean edge
fallback. These values estimate the accuracy cost of admission rejection
and are not end-to-end measurements. The global bound can reject
legitimate unauthenticated offloads and force edge fallback;
Section~\ref{sec:eval} quantifies this containment--utility trade-off.
These bounds concern
admitted request \emph{counts}, not a reserved GPU-time fraction, latency
guarantee, or minimum untrusted service share. Strict priority can starve
untrusted clients under trusted overload, which is outside the evaluated
threat model.

Accounting after the confidence gate charges actual cloud-demand events.
A pre-gate API limiter counts all requests, including local answers, so its
quota does not map one-to-one to cloud admissions when $q$ changes. This does
not make post-gate limiting a new primitive: a conventional aggregate limiter
placed at the same boundary can enforce the same count cap. Our contribution
is the composition, threat model, and measured-service evaluation of the
resulting isolation--rejection tradeoff.

The defense simulator uses two logical classes, finite capacity $K$ counting
waiting plus in-service requests, and bootstrap-sampled A100 service times.
The fixed case-study parameters are per-source depth $B_u=5$,
refill $R_u=0.69$/s, and $K_p=K/2=32$.

The two token bucket burst depths govern how many offload tokens can be accumulated, and
therefore how much short-term traffic can be admitted \cite{tokenbucket}. The per-source depth
$B_u{=}5$ lets one identity submit a burst of up to five offloads when its
bucket is full; the global depth $b_{\mathrm{gtb}}{=}20$ instead provides a
single bucket shared by all unauthenticated clients, allowing the untrusted
tier an aggregate initial burst of twenty. Summing the per-identity bound of
Eq.~\eqref{eq:bounds} over identities yields a quantity that grows with the
Sybil identity count, whereas the shared bucket's bound is
identity-independent: $B_u$ limits bursts from an individual source,
$b_{\mathrm{gtb}}$ the combined burst of the entire untrusted tier.

\textbf{Parameter provenance.}
Three of these values are fixed analytically from the workload definition
rather than searched: $R_u=1.5q_b=0.69$/s and
$r_{\mathrm{gtb}}=1.3(25)(0.46)=14.95$/s each give fixed headroom above the
honest per-identity and aggregate untrusted rates, and $K_p=K/2$ is a
symmetric capacity split; both refills are slack under honest load by
construction and bind only under a flood. The two burst depths ($B_u=5$,
$b_{\mathrm{gtb}}=20$) were instead checked on a disjoint clean-traffic
validation trace to keep authenticated false rejection low while leaving
headroom for flood onset (App.~\ref{app:provenance} gives the full
criterion and trace). No attacked trace, and no result in
\S\ref{sec:tradeoff}--\S\ref{sec:policy-tradeoff}, was used to select any
of these values, and we did not search the joint parameter space for a
best operating point; \S\ref{sec:policy-tradeoff} instead reports the full
global-bucket rate sweep, including rates that dominate the nominal
$14.95$/s setting on unauthenticated rejection. The \emph{concentrated and
Sybil flood construction} used throughout the defense evaluation is
described in \S\ref{sec:eval-setup}.

\medskip
\noindent\textit{Intuition: bound each identity, then bound the tier.}

\textbf{The question the defense asks.} Section~\ref{sec:queue} showed
that an attacker who drives $q_a\!\rightarrow\!1$ buys ${\approx}2.174\times$
cloud work per request and rides it through the knee at
$\varphi^{*}\!\approx\!0.50$. Bounded Escalation does not try to detect
$\delta$. It asks a different question of every offload request---\emph{who is
asking, and how much have they already been given?}---and answers it before
the request reaches the cloud.

\textbf{Step 1. Bound each identity.} A per-source token bucket
\cite{tokenbucket} caps how many offloads any single identity may have
admitted, so it contains concentrated attack traffic.
Splitting a fixed aggregate attack rate across sufficiently many identities
can place each identity below the per-source refill rate while leaving the
aggregate offered load unchanged. The per-identity bound of
Eq.~\eqref{eq:bounds} then holds for each identity, but their \emph{sum}
grows with the identity count (Eq.~\eqref{eq:sybilbound}): rate limiting
alone contains the flood it was designed for but collapses against the one
that matters.

\textbf{Step 2. Bound the tier, and the order of service.} Two mechanisms
recover authenticated service without reference to identity count. Protected
capacity admits untrusted requests only while occupancy is below
$K\!-\!K_p$, so untrusted floods cannot fill the buffer; non-preemptive
strict priority serves a waiting trusted request whenever the server frees,
though it does not interrupt a job already running. The optional global
untrusted bucket adds the identity-independent bound of
Eq.~\eqref{eq:gtbbound}, capping aggregate untrusted admissions no matter how
many identities share the load.

\section{Evaluation}
\label{sec:eval}

\subsection{Setup and Metrics}
\label{sec:eval-setup}

The public EdgeBoost artifact provides offline models and precomputed routing
outputs, but no request-arrival, cloud-serving, or queueing implementation
\cite{edgeboost2025,edgeboostrepo}. We extend it with attack generation and a
finite-capacity discrete-event queue simulator driven by measured A100
service-time samples; network, serialization, and host-to-device delays are
excluded. The artifact uses CIFAR-100, a $100$-class image-classification
benchmark~\cite{cifar100}. We use its shipped test records and model outputs
rather than retraining either the edge or cloud model. We distinguish hardware
measurements of inference latency and energy, measured-service simulations of
queueing and admission, and model-derived estimates of transport energy and
fallback accuracy.

The clean gate uses $8000$ shipped probability records. Primary per-input
margin attacks use $2000$ calibration/validation records and a fixed
$1000$-image direct-inference test subset, so named because attacks execute
the shipped edge model on images rather than replaying stored probabilities;
the queue itself is simulated throughout. Universal training uses a
disjoint $7000/1000/2000$ train/validation/test partition of the raw
test set (seed $42$), and is evaluated only on its held-out $2000$-image
test partition.
Application requests arrive as independent Poisson processes: each
principal offers one request per second with exponential inter-arrival times
drawn from the run seed, attacker and benign principals share the same
temporal model and differ only in offload probability, and no burst
correlation or arrival synchronization is modeled.

Table~\ref{tab:arrival} replays the same mean-rate and service-capacity
setting under three non-Poisson processes: phase-synchronized and
random-phase periodic arrivals, and batched arrivals ($B{=}4$ requests
every $4$\,s per principal, same long-run rate). At the saturated knee
($\varphi{=}0.5$, $\rho\approx1.00$), attacked benign p99 stays in
$1643$--$1737$\,ms under all four processes, so the high attacked tail is
not a Poisson artifact; the \emph{clean} baseline instead varies strongly
with temporal structure ($133$--$757$\,ms), so the amplification ratio
ranges from $2.26\times$ to $12.8\times$ and the \benignPninetynineAmp{}
headline depends on the arrival model as well as the operating point (the
sub-knee, $\varphi{=}0.4$, shows the same pattern: $266$--$1062$\,ms). Under
BE$+$G the flood stays contained under all four processes, though
synchronized arrivals degrade aggregate benign p99 to $388$\,ms against
$100$\,ms under Poisson, since per-source buckets admit bursts
token-by-token; this BE$+$G column uses the fixed-application-volume
construction of this section, so it is not directly comparable to
Table~\ref{tab:defenses}'s separate cloud-work-matched Sybil stress test.
The multi-gate test, budget sweep, and preprocessing evaluation each use
different splits, seed counts, and clean offload/accuracy baselines
(detailed in App.~\ref{app:protocols}) and should not be compared as
repeated measurements of one population.

\paragraph{Limit of plug-in system accuracy.}
The exact accuracy of a confidence-gated system can be written as
\[
A_{\mathrm{sys}}
=
(1-q)A_{\mathrm{edge}\mid o=0}
+
qA_{\mathrm{cloud}\mid o=1},
\]
where both accuracies are conditional on the routing outcome. The commonly
used plug-in approximation replaces the second term with the cloud model's
aggregate clean accuracy,
\[
\widehat{A}_{\mathrm{plugin}}
=
(1-q)A_{\mathrm{edge}\mid o=0}
+
qA_{\mathrm{cloud}}.
\]
That substitution is generally invalid here. The event $o=1$ is determined
by the input-dependent confidence gate, so offloaded examples are not a
random sample of the dataset and may have a different difficulty
distribution. Moreover, under attack, the perturbation can transfer from
the edge gate to the cloud classifier, making the cloud's accuracy on
perturbed offloads different from both its aggregate clean accuracy and its
accuracy on clean offloads. Thus, the plug-in expression is not a measured
end-to-end accuracy; it is an estimate that assumes away routing selection
and attack transfer. We therefore run the cloud model on the clean and
perturbed versions of the same gate-selected inputs and report their
paired, same-population accuracies.

\textbf{Measured attacked-cloud accuracy on an independent checkpoint.}
The EfficientNetV2-L checkpoint distributed with the EdgeBoost artifact
does not reproduce the artifact's precomputed $90.81\%$ cloud accuracy
under any of the $27$ tested input pipelines (best: $83.4\%$). We therefore
use a separately reproducible, independently trained EfficientNetV2-L
CIFAR-100 checkpoint\footnote{Checkpoint and implementation:
hankyul2/EfficientNetV2-pytorch,
\url{https://github.com/hankyul2/EfficientNetV2-pytorch}; the repository
reports $91.9\%$ accuracy. File name, SHA-256, and preprocessing are
recorded in the released evaluation harness.} and re-execute the white-box
escalation attack ($\tau=0.72$, $\epsilon=8/255$, $40$ steps, seed $42$),
evaluating both the clean and perturbed versions of every attacked-offload
input on the A100. This protocol produces a $46.3\%$ clean offload rate,
close to the full-artifact clean rate of $45.99\%$ at the same
threshold, and the attack reaches $99.9\%$ offload.

The attack sends $999$ of $1000$ inputs to the cloud. Among the $999$
inputs offloaded under attack, the independent cloud model correctly
classifies $905/999$ ($90.6\%$) of their clean versions and $823/999$
($82.4\%$) of their perturbed versions. Holding the image population
fixed, the perturbation therefore reduces cloud accuracy by $8.2$
percentage points. The attack is designed to amplify cloud demand, and it
also causes measurable cross-model prediction damage.

\textbf{Populations and uncertainty.}
At the matched $\varphi=0.5$ operating point the $25$ benign principals
comprise $12$ authenticated and $13$ unauthenticated; all adversarial
principals are untrusted. Latency statistics concern admitted cloud
requests arriving after warm-up and include the sampled service time;
local answers and rejected offloads are excluded, with rejection
reported per offload attempt and malicious share over all service
starts. \emph{Aggregate benign} is the sample-weighted union of the two
benign classes; the ``all benign'' population additionally includes
reconstructed local answers and rejected-offload fallbacks
(\S\ref{sec:deadline}), whose accuracy is model-derived rather than
measured end-to-end.

\begin{figure*}[!t]
  \centering
  \includegraphics[width=0.7\textwidth]{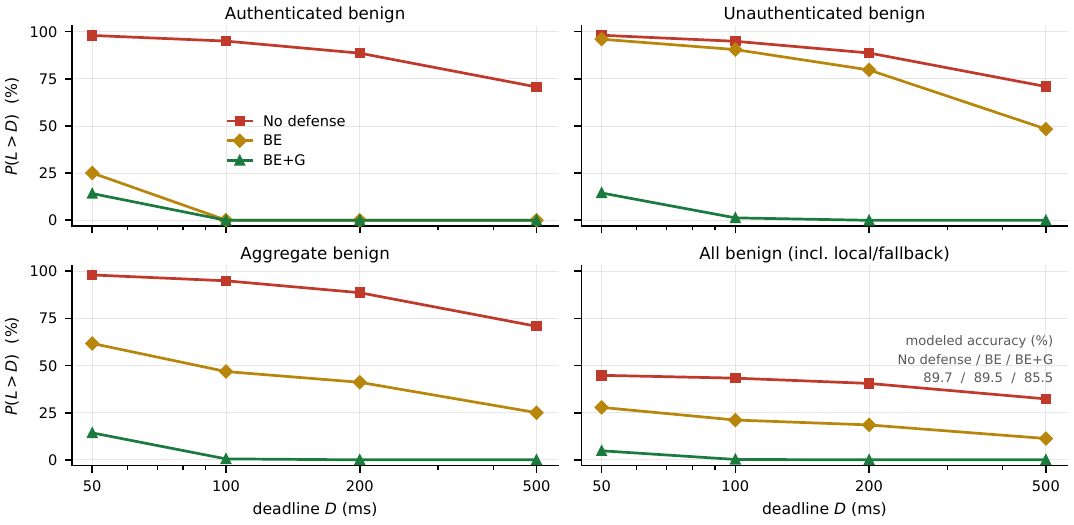}
  \caption{Deadline-miss probability $P(L>D)$ at $\varphi=0.5$ under the Sybil
  flood (12 pooled seeds, compute-and-queueing latency only). The first three
  panels concern admitted benign cloud requests; the fourth reconstructs all
  benign requests, counting local answers and rejected-offload fallbacks at the
  measured $1.30$\,ms edge latency. Under BE the unauthenticated tier retains a
  high miss rate ($90.5\%$ at $D{=}100$\,ms) while the authenticated tier does
  not ($0.1\%$), showing that unauthenticated clients bear the latency cost of
  isolation. The modeled accuracies annotated in the fourth panel are
  conditional model estimates, not hardware measurements: BE$+$G's near-zero
  miss rate is obtained partly by returning fallback predictions, at a cost of
  $4.2 (89.7-85.5)$ percentage points in modeled all-benign fallback accuracy.}
  \label{fig:deadline}
\end{figure*}

Table~\ref{tab:defenses} and Fig.~\ref{fig:deadline} use the same simulator and
the same seed range, $0$--$11$, and therefore should not be interpreted as
independent experimental evidence. Table~\ref{tab:defenses} reports means of
per-seed p99 values with $95\%$ $t$-intervals, whereas
Fig.~\ref{fig:deadline} pools latency samples to estimate deadline
probabilities; consequently, their uncertainty summaries need not coincide.
For example, the aggregate-benign p99 intervals are $1600\pm15$\,ms for BE
and $5262\pm131$\,ms for strict priority, while the authenticated-p99
half-widths are at most $4$\,ms.

A different uncertainty calculation applies to the budget and preprocessing
experiments. Treating each tested image as an independent trial with the same
success probability, we use two-sided exact $95\%$ Clopper--Pearson binomial
confidence intervals~\cite{clopper1934use}, which are more conservative than
the normal (Wald) approximation and therefore more reliable when success
rates lie near the $100\%$ boundary, as several do here
(App.~\ref{app:clopper} gives the intuition and a worked example). Under the
iid Bernoulli interpretation, the least favorable two-sided exact intervals
are $[99.44\%,100.00\%]$ for $999/1000$ budget-sweep successes and
$[99.28\%,99.98\%]$ for $998/1000$ preprocessing successes. Intervals are
computed per $1000$-image run; outcomes across attack seeds are not pooled
as independent trials and do not quantify variability across attack seeds,
model pairs, datasets, or deployment conditions.

\textbf{On-device measurement protocol.} The edge tier is a Jetson Orin Nano
Super running MobileNetV3-Small; the cloud is an NVIDIA A100-SXM4-80GB
running EfficientNetV2-L. Both use FP16 in a latency-oriented batch-one configuration with no
batching or concurrent kernels. After $50$ warm-up iterations, $1000$ batch-one inferences yield mean
latencies of $1.30$\,ms on the edge device and $27.4$\,ms on the cloud
GPU. Because the queue server models only cloud execution, its effective
serial service rate is
$\mu=1/\mathbb{E}[S_{\mathrm{cloud}}]\approx1/0.0274
=36.5$\,requests/s. The edge latency precedes cloud admission and is
included in per-request compute latency, but not in the cloud service time.

\textbf{Concentrated and Sybil flood construction.}
The defense evaluation of
\S\ref{sec:defense}--\S\ref{sec:policy-tradeoff}
(Tables~\ref{tab:defenses}, \ref{tab:revision-policies} and
Fig.~\ref{fig:sybil}) reuses the following attack-traffic construction
(full arithmetic in App.~\ref{app:flood}). We distinguish a
\emph{principal} (a real attacker entity or account, used for
authentication and per-source rate limiting) from an \emph{identity}
(a source presented to the admission system, which may not correspond
1:1 to a principal).

Of the $50$ total principals ($25$ benign and $25$ adversarial), a
\textbf{concentrated flood} presents the $25$ adversarial principals as
$25$ identities. Each identity sends one application request per second
and offloads with probability $1.0$, producing
$25\times1\times1.0=25$ adversarial offloads/s. A
\textbf{Sybil flood} splits each adversarial principal into $10$
identities, producing $250$ adversarial identities. Each still sends one
application request per second but offloads with probability $0.1$, so
the aggregate rate remains
$250\times1\times0.1=25$ adversarial offloads/s. The adversarial share of
visible identities therefore increases from $25/50=50\%$ to
$250/275=90.9\%$, while each Sybil identity individually appears as a
low-rate source. Both constructions thus impose the same aggregate adversarial cloud
input while distributing it differently across identities---the
dimension tested against the per-source token bucket
(\S\ref{sec:defense}). This differs from the primary escalation
experiments of \S\ref{sec:queue}, which hold total
\emph{application-level} request volume fixed at
$\Lambda=50$~req/s and let cloud demand vary with $q_a$. In the Sybil
construction, pre-gate application traffic instead rises to
$275$~req/s even though adversarial cloud-offload volume is matched to
the concentrated case. Table~\ref{tab:config} summarizes the remaining
experimental configuration.

\begin{table}[t]
\centering
\caption{Experimental configuration. Concentrated and Sybil floods offer
approximately $25$ adversarial offloads/s.}
\label{tab:config}
\footnotesize
\setlength{\tabcolsep}{5pt}
\begin{tabular}{@{}p{0.23\columnwidth}p{0.70\columnwidth}@{}}
\toprule
\textbf{Category} & \textbf{Configuration} \\
\midrule
Models/data & CIFAR-100; MobileNetV3-S edge; EfficientNetV2-L cloud \\
Gate & Temperature $T\approx2.13$; margin threshold $\tau=0.72$ \\
Hardware & Jetson Orin Nano Super; NVIDIA A100 \\
Service & Edge $1.30$\,ms; cloud $27.4$\,ms; $\mu\approx36.5$/s \\
Network delay & Excluded; compute-and-queueing latency only \\
Queue & Two-class M/G/1/$K$-style simulator, $K=64$; measured service \\
Traffic & $50$ pre-splitting principals; $25$ adversarial; Sybil expands identities at matched cloud input \\
Benign clients & $12$ authenticated / $13$ unauthenticated at $\varphi=0.5$ \\
Traces & $12$ seeds; defense and deadline traces use $100$\,s warm-up, queue-amplification and arrival traces use none; main/defense $3000$\,s, deadlines $2000$\,s \\
Per-source bucket & Depth $5$; refill $0.69$/s \\
Protected capacity & $K_p=32$ of $K=64$ positions \\
Global bucket & Depth $20$; nominal refill $14.95$/s \\
\bottomrule
\end{tabular}
\end{table}

\textbf{Expected calibration error.} ECE measures whether predicted
confidence agrees with empirical accuracy, not accuracy itself:
\[
\mathrm{ECE}=\sum_{m=1}^{M}\frac{|B_m|}{N}
\left|\operatorname{acc}(B_m)-\operatorname{conf}(B_m)\right|,
\]
where $\operatorname{acc}(B_m)$ is the fraction of correct top-1 predictions
in bin $B_m$ and $\operatorname{conf}(B_m)$ their mean predicted
confidence~\cite{guo2015calibration}. We compute ECE over all $N=1000$
edge predictions using $M=15$ equal-width confidence bins. ``Edge ECE''
refers only to predictions produced by the edge model before the routing
and cloud-inference stages; it is not a calibration measure for the combined
edge--cloud system. A low ECE does not imply high accuracy: a model can
become less accurate while remaining calibrated if its confidence decreases
by a similar amount.

\textbf{Clean gate characterization.} On shipped predictions, edge accuracy is \edgeAcc{} and cloud accuracy
\cloudAcc{}. Temperature scaling reduces ECE from \eceUncal{} to
\eceCal{} without changing top-1 labels. At $\tau=\opThreshold$,
offload is \cleanOffload{} and artifact-derived clean combined accuracy is
\cleanCombinedAcc{}, compared with \cloudAcc{} for cloud-only inference;
the cloud corrects
\cloudCorrection{} of offloaded edge errors.

The primary direct-inference subset has $46.3\%$ clean offload at the
deployed threshold, close to the full-artifact operating point of
$45.99\%$; the budget sweep and the preprocessing evaluation use the same
$1000$-image split. These subset-specific operating points are not
interchangeable
with the full-artifact operating point.

\subsection{Attack Effectiveness}
\label{sec:attack-results}
We first determine whether confidence-gate manipulation remains effective
beyond the strongest white-box setting. The experiments vary the attacker's
access to the victim model, whether the perturbation is optimized separately
for each input or reused across inputs, and the confidence signal used by
the gate; experiments with different data splits or gate thresholds are
separate evaluations rather than a direct ranking.

\textbf{Purpose and organization of Table~\ref{tab:attack}.}
Table~\ref{tab:attack} asks whether escalation remains effective when the
attacker has less knowledge or requires a reusable perturbation. All rows
target the deployed calibrated margin router. \emph{White-box margin} is
the strongest reference setting: a separate perturbation is optimized for
each input using full access to the victim edge model, calibrator, and gate.
\emph{Transfer} optimizes the perturbation on a MobileNetV2 surrogate and
then applies it to the victim without access to victim gradients.
\emph{Decision-only} observes only the binary local/offload outcome and
tests at most $30$ perturbed candidates after the clean query, without
using confidence scores or gradients. \emph{Universal} learns one
input-independent perturbation on a disjoint training split and reuses it
on all held-out inputs. Gate generality across max-softmax and entropy
routers is evaluated separately in Table~\ref{tab:cascade}. The ``Clean
offl.'' and ``Att.\ offl.'' columns report the fractions routed to the
cloud before and after attack, while ``Cloud amp.'' is their ratio
$q_a/q_b$.

Under the reproducible deployed-margin protocol, calibration-aware PGD
raises offload from $46.3\%$ to $99.9\%$ ($2.16\times$ cloud-demand
amplification), transfer raises it to $69.4\%$ ($1.50\times$), and
routing-decision-only search to $88.5\%$ ($1.91\times$). The white-box
attack also substantially disrupts the edge model's predictive behavior:
edge accuracy decreases from $73.1\%$ to $28.8\%$, edge ECE increases
from $3.142\%$ to $14.566\%$, and only $33.9\%$ of edge predictions are
preserved. Although the optimization objective targets the routing
margin, the resulting perturbations also alter class decisions and
degrade calibration. Edge ECE is computed over all edge-model outputs
before routing and cloud substitution; it characterizes the attacked
edge predictor, not the calibration of the final predictions served by
the combined edge--cloud system, which under attack are dominated by
cloud answers. In the independent-cloud evaluation of
\S\ref{sec:eval-setup}, the same-population cloud accuracy decreases
from $90.6\%$ on clean inputs to $82.4\%$ on their perturbed versions.
Thus, the attack both shifts nearly all requests to the shared cloud
service and causes measurable cross-model prediction damage.

Routing-decision-only search reaches $88.5\%$ offload with at most $31$
observations per image and succeeds on $78.58\%$ of initially local
inputs; transfer without victim gradients reaches $69.4\%$ with
$49.91\%$ initially-local success. A single reusable universal
perturbation, trained offline on a disjoint split, raises offload from
$45.85\%$ to $78.75\%$ on its held-out evaluation partition
($1.72\times$, $67.2\%$ initially-local success). Thus lower-knowledge
and reusable attacks also substantially increase cloud demand, though
less than white-box per-input optimization.

\begin{table}[t]
\centering
\caption{Escalation attacks against the calibrated margin router.
White-box, transfer, and decision-only use the same $N=1000$
direct-inference subset; universal uses a separate held-out $N=2000$
partition. Local success is measured only over inputs initially
processed locally.}
\label{tab:attack}
\footnotesize
\setlength{\tabcolsep}{2pt}
\renewcommand{\arraystretch}{1.05}
\begin{tabular}{@{}lcccc@{}}
\toprule
\textbf{Attack} &
\shortstack{\textbf{Clean}\\\textbf{offl.}} &
\shortstack{\textbf{Attacked}\\\textbf{offl.}} &
\shortstack{\textbf{Cloud}\\\textbf{amp.}} &
\shortstack{\textbf{Local}\\\textbf{succ.}} \\
\midrule
WB margin
  & 46.3\%  & 99.9\%  & 2.16$\times$ & 99.81\% \\
Transfer
  & 46.3\%  & 69.4\%  & 1.50$\times$ & 49.91\% \\
Decision-only
  & 46.3\%  & 88.5\%  & 1.91$\times$ & 78.58\% \\
Universal
  & 45.85\% & 78.75\% & 1.72$\times$ & 67.22\% \\
\bottomrule
\end{tabular}

\vspace{1pt}
\parbox{\columnwidth}{\scriptsize
WB: white-box. Cloud amp. is the attacked-to-clean offload ratio.
Decision-only uses one clean observation and up to 30 candidate queries
($\leq31$ total). These search queries are excluded from the online queue
workload. The reproducible protocol uses evaluation mode,
$\tau=0.72$, $T=2.1272$, records $2000{:}3000$, resize/crop to
$224$, and ImageNet normalization. Gate generality is reported separately
in Table~\ref{tab:cascade}.
}
\end{table}

\textbf{Confidence gates.}
A confidence gate converts the edge model's calibrated probability vector
$p(x)$ into a binary local/offload decision. We evaluate three common gate
signals. The \emph{margin gate} measures the difference between the largest
and second-largest probabilities,
$m_T(x)=p^{(1)}(x)-p^{(2)}(x)$, and offloads when $m_T(x)\leq\tau_m$; a
small top-two gap means the model has difficulty choosing between its two
leading classes. The \emph{max-softmax gate} uses only the largest predicted
probability, $s(x)=\max_j p_j(x)$, and offloads when $s(x)\leq\tau_s$;
even if one class ranks first, a low maximum probability indicates weak
confidence in that choice. The \emph{entropy gate} uses
\[
H(p(x))=-\sum_j p_j(x)\log p_j(x)
\]
and offloads when $H(p(x))\geq\tau_H$; high entropy means probability mass
is spread across several classes rather than concentrated on one prediction.
Accordingly, PGD seeks to decrease the margin for the first gate, decrease
the maximum probability for the second, and increase entropy for the third.
A single distribution shift can move all three signals: changing
$p(x)=[0.60,0.30,0.10]$ to $p(x')=[0.40,0.35,0.25]$ shrinks the margin,
lowers the maximum probability, and raises entropy without changing the top
class. The three thresholds are calibrated separately on a disjoint
validation set to produce approximately the same clean offload rate, so
their numerical values are not directly comparable.

\textbf{Multi-gate generality.}
Separate image-space PGD attacks against the three gates reach
$99.9$--$100\%$ offload from approximately $45\%$ clean offload
(Table~\ref{tab:cascade}). At least $99.8\%$ of initially-local inputs cross
the gate, attacked/clean offload is at least $2.20\times$, and edge accuracy
falls to $27$--$28\%$. This experiment tests whether the escalation attack
extends beyond one confidence definition; it does not establish robustness
or vulnerability for every possible routing rule. The margin row uses its
own frozen validation-calibrated threshold and therefore differs slightly
from the deployed-$\tau$ test in Table~\ref{tab:attack}: $45.40\%$
versus $46.3\%$ clean offload, and $2.20\times$ versus $2.16\times$
amplification.

\begin{table*}[t]
\centering
\caption{Multi-gate PGD evaluation with $\epsilon=8/255$, $T\approx2.13$,
and $N=1000$ test images. The top-two-margin and maximum-softmax gates
offload when their confidence signals fall below their respective
thresholds, whereas the entropy gate offloads when predictive entropy
exceeds its threshold. Gate-specific thresholds are selected on a disjoint
validation split to target approximately $46\%$ clean offload and then
frozen for test evaluation; they are not the deployed margin threshold
$\tau=0.72$. ``Success'' denotes an initially-local
input that crosses the corresponding gate, and cloud amplification is the
attacked-to-clean offload ratio.}
\label{tab:cascade}
\small
\setlength{\tabcolsep}{6pt}
\begin{tabular}{@{}lcccccc@{}}
\toprule
\textbf{Confidence gate} & \textbf{$\tau$} & \textbf{Clean off.} & \textbf{Att.\ off.} & \textbf{Success (init.\ local)} & \textbf{Cloud amp.} & \textbf{Edge acc (att.)} \\
\midrule
Top-two margin & 0.702 & 454/1000 ($45.40\%$) & 1000/1000 ($100.00\%$) & 546/546 ($100.00\%$) & 2.20$\times$ & 28.0\% \\
Maximum softmax & 0.794 & 447/1000 ($44.70\%$) & 999/1000 ($99.90\%$) & 552/553 ($99.82\%$) & 2.23$\times$ & 27.3\% \\
Predictive entropy & 0.810 & 447/1000 ($44.70\%$) & 1000/1000 ($100.00\%$) & 553/553 ($100.00\%$) & 2.24$\times$ & 28.1\% \\
\bottomrule
\end{tabular}
\end{table*}

\textbf{Natural-input baseline.}\label{sec:selection}
This baseline asks whether high offload can be achieved without
perturbing any pixel, by selecting naturally occurring images. Among
$8000$ stored records, $45.99\%$ already offload at $\tau=0.72$; an
attacker with offline routing information can simply select such
records and obtain $100\%$ offload. A weaker attacker without this
information who draws $k$ independent candidates and submits whichever
offloads achieves probability $1-(1-q_b)^k$; for $k{=}8$, simulation
gives $99.23\%$ across eight seeds. These are offline selection queries
over a fixed legitimate pool, not extra online requests or perturbations
of a prescribed image; this baseline requires freedom to select the
submitted image and therefore does not replace the prescribed-input
perturbation threat model of Section~\ref{sec:attacks}.

\subsection{Queue Amplification}
\label{sec:amp}
A high attack success rate at the confidence gate does not by itself
establish a serious availability impact. This subsection connects the
attack-induced offload probability to offered cloud load and separates two
effects: the direct additional computation caused by executing the cloud
model, and the tail-latency amplification caused by queueing near the
service-capacity boundary. Fixed-volume, service-model, and arrival-process
comparisons identify which part of the observed amplification is specific to
the evaluated operating conditions.

\begin{figure}[t]
\centering
\includegraphics[width=0.8\columnwidth]{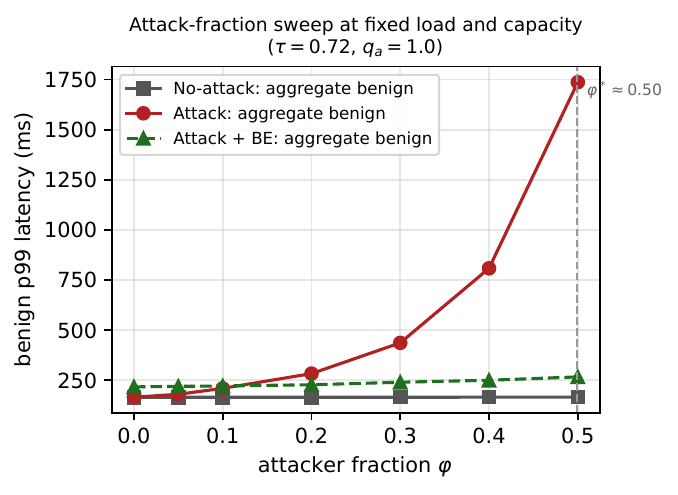}
\caption{Admitted aggregate-benign p99 versus attacker fraction at
fixed application volume and capacity ($\Lambda=50$\,req/s,
$\mu\approx36.5$/s, $K{=}64$, $12$ seeds, measured A100 service; network
and serialization delay excluded). All three curves report the same
population: the no-attack baseline and the escalation attack use no
defense, and the defense curve applies Bounded Escalation to the attack
case. The dashed line marks the analytic offered-load critical point
$\varphi^\ast{\approx}0.50$ (Eq.~\eqref{eq:critical}), computed for the
undefended system. Rejected offloads are excluded from every curve.}
\label{fig:p99}
\end{figure}

\textbf{Attack impact at fixed request volume.}
The controlled comparison holds the number of principals, their arrival
processes, and the aggregate application rate $\Lambda=50$\,req/s fixed,
and changes only whether the attacker-controlled fraction $\varphi$
submits clean or perturbed inputs. The attack-induced increase in offered
cloud load is then
$\Delta\lambda_c=\Lambda\varphi(q_a-q_b)$, attributable to offload
manipulation rather than request volume. Across \cfSeeds{} seeds, the
no-attack p99 stays at \cfPnnClean{}\,ms ($\rho=\cfRhoClean{}$) for
every $\varphi$; under attack, benign p99 rises from \cfPnnClean{}\,ms at
$\varphi=0$ to $834$\,ms at $\varphi=0.4$ ($\rho=0.93$) and
$1738$\,ms at $\varphi=0.5$ ($\rho=1.00$), where $P(L{>}100\,\mathrm{ms})$
reaches $95.4\%$. Above the knee the system is saturated: at
$\varphi\in\{0.6,0.8\}$, p99 plateaus near $1755$--$1758$\,ms and
rejection, not further delay, absorbs the excess load
(Fig.~\ref{fig:p99}).

\begin{figure*}[hbtp]
\centering
\includegraphics[width=0.9\textwidth]{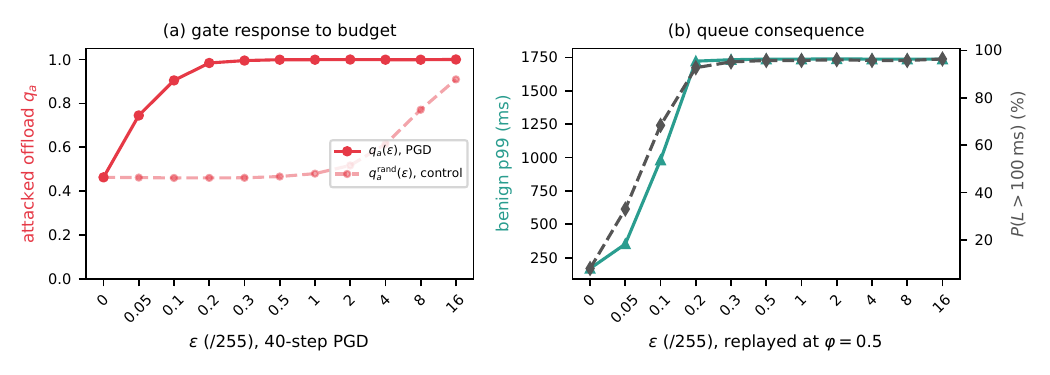}
\caption{Budget sweep: (a) attacked offload under PGD and the random-sign
control;
(b) admitted aggregate-benign p99 and $100$\,ms miss probability under
$\varphi=0.5$ replay. Near-complete offload at $0.5/255$ puts the replay at
the capacity boundary, $\rho\approx1.00$.}
\label{fig:epssweep}
\end{figure*}

\textbf{Perturbation budget.}
Sweeping $\epsilon\in\{0,0.05,0.1,0.2,0.3,0.5,1,2,4,8,16\}/255$ on
$N=\epsN{}$ test images (three attack seeds, $40$ steps,
$\alpha=\min(1/255,\epsilon/4)$, $\tau=0.72$), PGD leaves the clean
rate $q_b=\epsQb{}$ immediately and reaches near-complete offload
($q_a=0.999$--$1.000$) by $\epsMinSat{}/255$, while the random
$\pm\epsilon$ sign-noise control stays near the clean rate across that
entire range and only reaches $0.909$ at $16/255$---a budget roughly
$32\times$ larger (Fig.~\ref{fig:epssweep}a). The separation between the
two curves distinguishes a gate collapsed by optimization from one
collapsed by an oversized perturbation ball; saved-tensor checks confirm
every budget stays within its pixel-space bound. Replaying each measured
$q_a(\epsilon)$ at $\varphi=0.5$ through the measured-service queue
(Fig.~\ref{fig:epssweep}b), admitted aggregate-benign p99 and the
$100$\,ms miss probability rise together through $0.05$--$0.3/255$
($352$--$1732$\,ms) and flatten near $1737$\,ms once $q_a$ saturates.
That plateau is an artifact of the replay input, not a robustness
property: every budget at or above $\epsMinSat{}/255$ supplies
$q_a=0.999$--$1.000$ to the simulator, coinciding with the analytic
boundary of Eq.~\eqref{eq:critical}. A convergence check at
$\epsilon\in\{2,4\}/255$ confirms the saturation is not an artifact of
insufficient PGD steps: one step already reaches $q_a=\convOneStepTwo{}$
and $\convOneStepFour{}$, and five steps reach \convFiveStep{}.

\textbf{Service-model sensitivity.}
Across exponential (M/M/1/$K$), deterministic (M/D/1/$K$), and measured
(M/G/1/$K$) service-time distributions with the Poisson arrival process
and $K=64$ fixed, the queueing knee remains between $\varphi=0.40$ and
$0.50$ (one $\Delta\varphi=0.1$ sweep interval), while the maximum benign
p99 ranges from approximately $1.75$ to $2.13$\,s: the qualitative
transition persists across service models, while its tail magnitude
remains configuration-dependent.

At the full-artifact queue operating point, forcing $q_a$ from
$q_b=0.4599$ toward one increases cloud demand per attacker request by at most
$1/q_b\approx$\cloudWorkAmp{}. The compute-only mean latency
ratio is \latencyAmp{}. With $50$ requests/s, serial batch-one service, and
$K=64$, undefended aggregate-benign p99 rises from
\cleanBenignPninetynine{} to \attackBenignPninetynine{}
(\benignPninetynineAmp{}) near $\varphi^\ast\approx\phaseTransition$.
This disproportionate tail increase depends on starting near capacity; the
fixed-volume comparison above isolates it from increased request volume.

\textbf{Direct per-request cost amplification.}
Before queueing amplification, changing the routing decision has a direct
per-request cost. Let $q$ denote the offloaded fraction of offered requests.
Every request executes the edge model, and an offloaded request additionally
executes the cloud model, so the mean compute-only inference time per offered
request is
\[
\bar L(q)=L_{\mathrm{edge}}+qL_{\mathrm{cloud}},
\]
with measured means $L_{\mathrm{edge}}=1.30$\,ms and
$L_{\mathrm{cloud}}=27.4$\,ms. Substituting $q_b=0.4599$ and $q_a=0.999$
gives \latencyClean{} and \latencyAttacked{}, a \latencyAmp{} increase.
This analytical average replaces the cloud service time by its measured
mean, $L_{\mathrm{cloud}}=\mathbb{E}[S_{\mathrm{cloud}}]$. The queue
simulator instead samples an individual service time $S_{\mathrm{cloud},i}$
from the complete empirical distribution for each admitted offload $i$ and
computes
\[
L_i^{\mathrm{cloud}}=W_i+S_{\mathrm{cloud},i},
\]
where $W_i$ is that request's simulated queueing delay. Thus, the p99 and
deadline results in Table~\ref{tab:defenses} and Fig.~\ref{fig:deadline}
preserve the measured request-to-request service-time variation. The gap
between the approximately twofold direct compute increase and the
$10.46\times$ p99 increase quantifies the additional amplification caused
by queueing near the cloud-capacity boundary.

Neither calculation includes network transmission, serialization, or
protocol overhead. Adding a fixed network delay $d_{\mathrm{net}}$ to each
offload would increase its end-to-end latency and could increase the
deadline-miss probability. Because arrivals are externally fixed and do not
depend on response time, however, this additive delay would not change the
offered cloud arrival rate $\lambda_c$, service capacity $\mu$, or the
queue-stability boundary $\lambda_c=\mu$.

\subsection{Defense Effectiveness}
\label{sec:tradeoff}
We next evaluate whether Bounded Escalation protects benign traffic after
an attacker has manipulated the confidence gate. A useful defense must be
judged along three separate dimensions: latency isolation for authenticated
clients, containment of aggregate untrusted cloud work, and collateral cost
to legitimate unauthenticated clients. We compare the component mechanisms
and their composition under concentrated and Sybil attacks, then measure
their overhead under clean load.

Table~\ref{tab:defenses} compares six configurations at matched attack volume.
Per-source limiting lowers authenticated p99 to \tbConcPninetynine{} under
the concentrated flood but leaves \tbSybilPninetynine{} under Sybil.
Protected capacity alone eliminates authenticated rejection yet leaves
approximately $886$\,ms p99: admission headroom does not control service order.

Strict priority alone already satisfies the evaluated authenticated criteria:
$77$\,ms p99 under both floods and less than $1\%$ rejection.
BE gives comparable $75$/$76$\,ms p99. However, strict priority shifts delay
onto unauthenticated traffic: Sybil aggregate-benign p99 is
$5262\pm131$\,ms, worse than no defense at $1738\pm1$\,ms.
BE reduces this tail to $1600\pm15$\,ms; BE$+$G reduces it to
$85\pm1$\,ms. All intervals are $95\%$ CIs for mean per-seed p99.

The global bucket adds an aggregate admission bound, reducing malicious cloud
share from \beMalSybil{} with BE to \gtbMalSybil{} with BE$+$G.
Authenticated rejection remains at most \gtbAuthLossMax{}, but legitimate
unauthenticated offload rejection rises from \beUnauthLossSybil{} to
\gtbUnauthLossSybil{}. Under clean routing-conditional correctness assumptions,
modeled all-benign fallback accuracy falls from
\benignAccNoDef{} without defense to \benignAccGtbSybil{} with BE$+$G,
a \benignAccDropSybil{}-point decrease; this is not measured accuracy on
defended hardware or an unauthenticated-only estimate.

These results show that per-source admission and priority alone do not
bound aggregate untrusted work under Sybil scaling; an identity-independent
aggregate admission mechanism is required for that objective. Authenticated
isolation, aggregate-work containment, and unauthenticated utility are
distinct objectives. BE$+$G is suitable only if best-effort unauthenticated
service is acceptable. Strict priority provides much of the authenticated
benefit without the global cap, while leaving a severe untrusted queue. Because an ordinary limiter at the same admission boundary
can also enforce an aggregate count bound, Section~\ref{sec:policy-tradeoff}
separately compares BE$+$G with a global-bucket-only FIFO baseline;
Table~\ref{tab:defenses} evaluates this composition rather than establishing
novelty over all admission policies.

\textbf{Clean-load overhead.}
Under clean mixed traffic, BE accelerates authenticated p99 from $166$ to
\cleanAuthPninetynine{} through strict-priority service order, but
this is a class-redistribution effect rather than a system-wide
improvement: unauthenticated-benign p99 rises to $260$\,ms and
aggregate-benign p99 increases from $166$ to $217$\,ms, so deployments
with meaningful anonymous
traffic bear the displacement. The full stack also rejects
\cleanAuthLossBE{} of authenticated and $1.23\%$ of unauthenticated
offloads even under clean load. BE and BE$+$G report identical clean
outcomes, indicating the global bucket is slack in this workload, not
that it is cost-free for arbitrary legitimate demand.

\begin{table}[t]
\centering
\caption{Arrival-process sensitivity under the fixed-application-volume
concentrated attack at $\varphi{=}0.5$: means of per-seed benign p99 over
$12$ seeds with pointwise $95\%$ $t$-intervals, measured service times,
$K{=}64$, $3000$\,s horizon, no warm-up (matching the primary
queue-amplification traces; the $100$\,s warm-up of Table~\ref{tab:config}
applies to the defense and deadline traces). Clean and attack columns use no
defense; the BE$+$G column applies Bounded Escalation$+$G to the same
fixed-volume workload and is not the cloud-work-matched Sybil stress of
Table~\ref{tab:defenses}. ``C.'' stands for Clean, and ``A.'' stands for Attack. ``Ampl.''\ is the ratio of the displayed integer mean p99 values, matching the headline's display convention. The last column is the no-defense attack p99 at $\varphi{=}0.4$
($\rho\approx0.93$). Batched arrivals use random per-principal phase.}
\label{tab:arrival}
\scriptsize
\setlength{\tabcolsep}{2pt}
\begin{tabular}{@{}lccccc@{}}
\toprule
\textbf{Arrival process} & \textbf{C. p99(ms)} & \textbf{A. p99(ms)} & \textbf{Ampl.} & \textbf{BE$+$G p99 (ms)} & \textbf{A. p99(ms)} \\
\midrule
Poisson (this paper) & 166{\scriptsize $\pm$3} & 1737{\scriptsize $\pm$2} & 10.46$\times$ & 100{\scriptsize $\pm$1} & 809{\scriptsize $\pm$58} \\
Periodic, synchronized & 757{\scriptsize $\pm$6} & 1712{\scriptsize $\pm$7} & 2.26$\times$ & 388{\scriptsize $\pm$4} & 1062{\scriptsize $\pm$3} \\
Periodic, random phase & 133{\scriptsize $\pm$10} & 1702{\scriptsize $\pm$10} & 12.8$\times$ & 154{\scriptsize $\pm$28} & 266{\scriptsize $\pm$20} \\
Batched ($B{=}4$) & 302{\scriptsize $\pm$31} & 1643{\scriptsize $\pm$40} & 5.44$\times$ & 224{\scriptsize $\pm$38} & 632{\scriptsize $\pm$90} \\
\bottomrule
\end{tabular}
\end{table}

\begin{table}[t]
\centering
\caption{Matched concentrated and Sybil floods (construction defined in
\S\ref{sec:eval}): measured-service queue,
$K=64$, $12$ seeds. Rejection denominators are class-specific offload
attempts; malicious share uses cloud service starts. Aggregate p99 includes
admitted benign requests only. Its $95\%$ CI is shown; authenticated-p99
half-widths are $\le4$\,ms. PSB: per-source bucket; PC: protected capacity;
SP: strict priority; BE: Bounded Escalation; G: global untrusted bucket.}
\label{tab:defenses}
\footnotesize
\setlength{\tabcolsep}{2pt}
\begin{tabular}{@{}lrrrrrrr@{}}
\toprule
& \multicolumn{4}{c}{\textbf{Auth.\ protection}} & \multicolumn{3}{c}{\textbf{System-wide (Sybil)}} \\
\cmidrule(lr){2-5}\cmidrule(lr){6-8}
& \multicolumn{2}{c}{p99 (ms)} & \multicolumn{2}{c}{rej.} & Agg.-benign p99 & Mal. & Unauth. \\
\cmidrule(lr){2-3}\cmidrule(lr){4-5}
\textbf{Def.} & Con. & Syb. & Con. & Syb. & (ms, 95\% CI) & share & rej. \\
\midrule
None & 1736 & 1738 & 0.81\% & 0.84\% & 1738{\scriptsize$\pm$1} & 68\% & 0.8\% \\
PSB & 252 & 1730 & 1.27\% & 1.78\% & 1729{\scriptsize$\pm$4} & 69\% & 1.8\% \\
PC & 886 & 886 & 0.00\% & 0.00\% & 878{\scriptsize$\pm$1} & 68\% & 1.9\% \\
SP & 77 & 77 & 0.81\% & 0.84\% & 5262{\scriptsize$\pm$131} & 68\% & 0.8\% \\
BE & 75 & 77 & 1.28\% & 1.25\% & 1600{\scriptsize$\pm$15} & 69\% & 2.8\% \\
BE$+$G & 72 & 72 & 1.29\% & 1.28\% & 85{\scriptsize$\pm$1} & 59\% & 52.1\% \\
\bottomrule
\end{tabular}

\end{table}

\subsection{Adaptive and Sensitivity Tests}
\label{sec:preddef}

Admission control bounds the cloud work a successful escalation attack can
create, but it does not address the confidence gate's own vulnerability:
the gate can still be fooled into escalating an input it would otherwise
have answered locally. We separately evaluate input preprocessing, a
prediction-side defense that instead attempts to remove or weaken the
perturbation before edge inference, restoring the routing decision the
gate would have made on the unperturbed input. Unlike admission control,
whose effectiveness does not depend on detecting or reversing the
perturbation, preprocessing targets the perturbation directly. The
evaluation distinguishes apparent protection against an oblivious attacker
from protection against an adaptive attacker, and also reports the
transformations' effects on clean accuracy and offload. Preprocessing
transforms an input before it reaches the edge model and confidence gate,
\[
x+\delta \xrightarrow{\text{preprocessing }g} g(x+\delta)
\xrightarrow{\text{edge model}} \text{confidence gate},
\]
so that the malicious offload rate
\[
q_a^{\mathrm{def}}=\Pr\!\left[o\bigl(g(x+\delta)\bigr)=1\right]
\]
falls. We test whether such preprocessing reduces attacked offload under
an oblivious attacker, whether the apparent benefit survives adaptive EOT
or BPDA optimization, and what cost the transformation imposes on clean
edge accuracy and clean offload. These baselines do not bound admitted
cloud work by construction; they test whether prediction-side input
transformation can prevent the gate manipulation that creates that work.

\begin{figure*}[!t]
  \centering
  \includegraphics[width=0.8\textwidth]{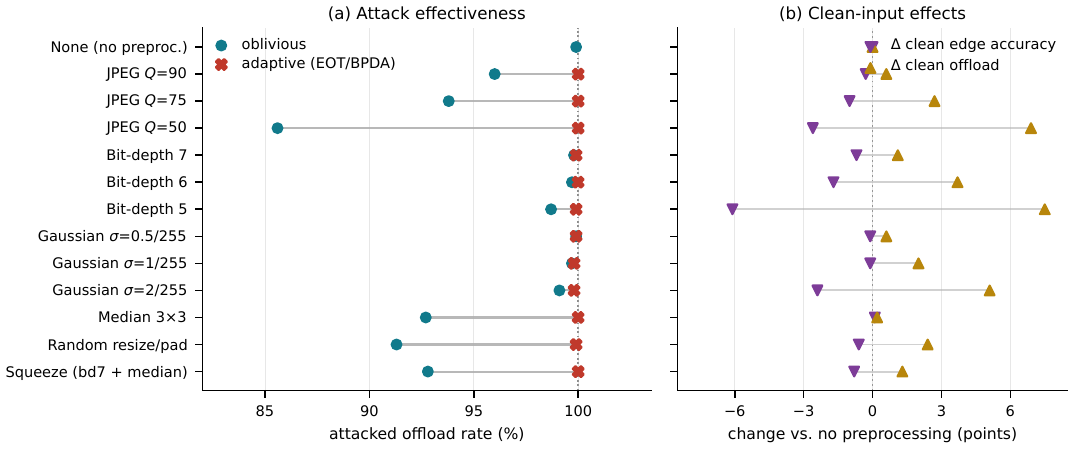}
  \caption{Preprocessing evaluation on the complete $N{=}1000$-image
  artifact split
  ($\epsilon=8/255$, 40-step PGD, one primary preprocessing seed). (a)
  Oblivious attacks omit the transform; adaptive attacks incorporate it
  through EOT (Gaussian noise, resize/pad) or a straight-through
  backward-pass differentiable approximation (BPDA) gradient (JPEG,
  bit-depth, median). Every evaluated preprocessing setting
  returns to at least $99.8\%$ attacked offload under adaptive optimization
  (six of twelve reach exactly $100\%$), so the apparent robustness to the oblivious
  attack does not survive adaptive evaluation and is consistent with
  gradient obfuscation rather than a robust reduction of the escalation
  surface. Panel~(b) evaluates the cost of preprocessing on clean,
  unattacked inputs. Each marker reports the change in edge accuracy or
  clean offload rate relative to no preprocessing, whose baseline values
  are \(73.1\%\) and \(46.3\%\), respectively. Negative accuracy changes
  indicate that preprocessing causes more edge prediction errors, while
  positive offload changes indicate that it sends more benign requests to
  the cloud. Panel~(a) reports attacked offload over all $1000$ test inputs
  because this quantity directly determines offered cloud load.
  Initially-local success, reported in the text, instead conditions on
  inputs that were answered locally by the unprocessed clean gate and
  measures how often the attack changes their routing decision.}
  \label{fig:preproc}
\end{figure*}

We evaluate \pdNumDefenses{} training-free preprocessing settings:
JPEG ($q\in\{90,75,50\}$), bit-depth reduction ($7$, $6$, $5$ bits/channel),
Gaussian noise ($\sigma\in\{0.5,1,2\}/255$), a $3\times3$ median filter,
randomized resize/pad, and bit-depth-plus-median feature squeezing.
Each is applied before the edge gate on the complete $N=\pdN{}$-image
artifact split, with one primary attack seed ($42$; seeds $43$ and $44$ replicate the
adaptive results within $0.2$ percentage points), $\epsilon=8/255$, and $40$
PGD steps.

The oblivious attack is optimized without preprocessing. Adaptive attacks
include the transform, using EOT ($8$ samples/step) for Gaussian noise and
resize/pad, and an identity BPDA gradient for JPEG, bit-depth, and median
filtering (both techniques are defined in Section~\ref{sec:related-adaptive}).
The initially-local success denominator is defined by the unprocessed clean
gate on this subset.

Across the twelve settings, the oblivious attack's apparent effectiveness
ranges from JPEG $Q=50$ (offload lowered to $85.6\%$, the strongest
oblivious result, with $77\%$ initially-local success) to feature
squeezing and randomized resize/pad (offload only lowered to
$91$--$93\%$, at $88\%$ initially-local success); Fig.~\ref{fig:preproc}(a)
plots oblivious and adaptive offload for all twelve settings. Every
tested adaptive attack raises offload back to at least $99.8\%$, and all
but the weakest bit-depth and Gaussian settings reach $100\%$
(Fig.~\ref{fig:preproc}). Adaptive optimization defeats all tested preprocessing
defenses on the full split. Physical camera-pipeline robustness remains
outside this evaluation.

These adaptive results challenge DefQ's conjecture that
gradient-obfuscation countermeasures would survive adaptive
attacks~\cite{qiu2023defq} (Section~\ref{sec:related-adaptive}): under BPDA and EOT,
all \pdNumDefenses{} tested preprocessing settings raise offload back to at
least $99.8\%$. %

Preprocessing affects not only attacker traffic but also clean unattacked images: the strongest
preprocessing settings reduce clean edge accuracy by up to 6.1 points and
raise clean offload by up to 7.5 points (Fig.~\ref{fig:preproc}), because degrading the
image also makes the edge model less confident about entirely benign
inputs, pushing more of them to the cloud. No setting improves clean edge
accuracy on the full test split---the tested preprocessing defenses carry a
real cost on normal traffic with no accuracy benefit in return. We also
omit clean combined-accuracy comparisons here, since such an
estimate would need to assume that the cloud model remains correct on
preprocessed images, an assumption we have not verified; we report no
number rather than one built on an untested assumption.

Given that preprocessing defenses do not work, \emph{admission-side controls} takes a different approach:
rather than attempting to restore the original routing decision, they limit
the cloud work admitted after an offload request reaches the admission
boundary. Per-source budgets limit individual identities, while the optional
global bucket bounds aggregate untrusted admissions independently of identity
count. 

\textbf{Defense-aware traffic attacks.}
\label{sec:adaptive}

The preceding experiments use a fixed traffic construction, whereas an
attacker who knows the admission policy may reorganize its requests to
exploit the policy's operating assumptions. This section asks whether such
an attacker can weaken containment by adapting the organization of its
traffic rather than the image perturbation itself: unlike the adaptive
attacks of Section~\ref{sec:preddef}, which incorporate input preprocessing
into the optimization, the strategies considered here are \emph{defense-aware
traffic attacks} that leave the underlying image attack unchanged and vary
how its offload volume is distributed across identities and time. An identity that offloads at $1$/s against a per-source
refill of $R_u=0.69$/s is limited by the bucket; spreading the same volume
over ten identities at $0.1$/s each keeps every identity below $R_u$ while
preserving the aggregate $10\,(0.1)=1$/s. Rate shaping instead holds each
identity near $0.99R_u$, and duty cycling concentrates requests into
alternating active and inactive intervals to exploit accumulated burst
tokens. We evaluate whether these strategies can increase malicious cloud
share or benign tail latency beyond the levels maintained by BE$+$G, and
report the rejection cost imposed on legitimate unauthenticated traffic.

\begin{figure*}[!t]
  \centering
  \includegraphics[width=0.8\textwidth]{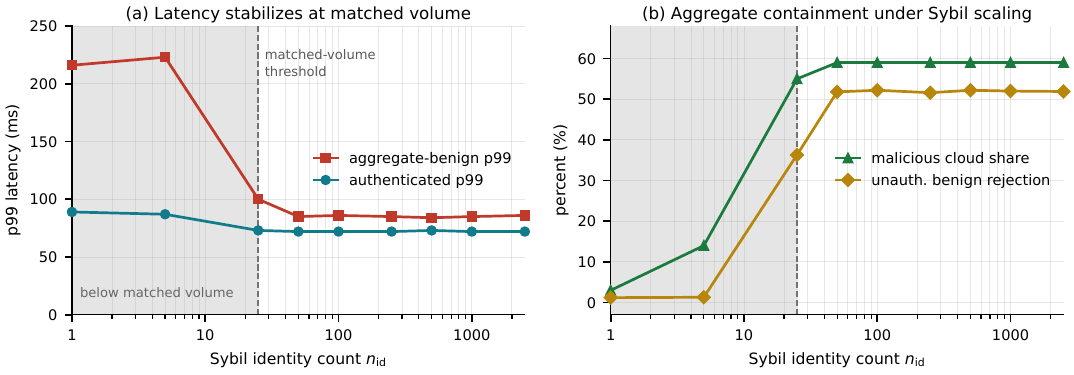}
  \caption{Sybil identity-count sweep against Bounded Escalation$+$G, using
the concentrated/Sybil flood construction of \S\ref{sec:eval} (12 seeds,
measured-service queue, $K=64$). The shaded region left of the dashed
matched-volume threshold contains configurations with $n_{\mathrm{id}}<25$,
which cannot reach the matched $\approx 25$ offloads/s and are shown as
low-volume references. (a) Once matched volume is reached, authenticated p99
remains nearly constant, while aggregate-benign p99 stabilizes by
$n_{\mathrm{id}}=50$. Neither metric grows with additional identities because
protected capacity and priority do not reference identity count. (b) The global bucket holds malicious cloud share at $\le 59\%$ across more than
  three orders of magnitude in identity count. The containment follows from
  protected capacity, priority scheduling, and the identity-independent
  global bucket, without anomaly detection. Rate-shaped and duty-cycled
  attackers are reported separately in Table~\ref{tab:adaptive}.}
  \label{fig:sybil}
\end{figure*}

We test parameter-aware traffic shaping against BE$+$G
(\aaSeeds{} seeds; Fig.~\ref{fig:sybil} and Table~\ref{tab:adaptive}).
Sybil scaling uses $1$--$2500$ \emph{identities} in the sense of
\S\ref{sec:eval-setup} (the $25$ adversarial principals of the
matched-flood comparison redistribute their aggregate offload volume
across $n_{\mathrm{id}}$ identities, so $n_{\mathrm{id}}=100$ means $100$
identities total, not $25\times100$), offering approximately $25$
offloads/s once $n_{\mathrm{id}}\ge25$; the $1$- and $5$-identity
configurations are lower-volume references. Rate-shaping configurations
set per-identity rates to $0.80$--$1.20R_u$. Duty cycling alternates
perturbed and clean traffic with a $5$\,s period (duty $d$ means the
attack is active for $5d$\,s of each period), probing whether burst timing
can extract more admitted cloud work than steady sending within the same
token budgets. These constructions are separate from the matched-flood
comparison; equal nominal identity counts do not establish
identical workloads.

Across the tested configurations, authenticated p99 remains below $90$\,ms
and malicious cloud share at most \aaMalShareMax{}.
Unauthenticated legitimate rejection reaches approximately
\aaUnauthRejMax{}. These empirical outcomes do not follow from the token
count bound alone, and do not cover joint optimization over images,
identities, and arrival times.

Fig.~\ref{fig:sybil} and Table~\ref{tab:adaptive} probe one question
along two attacker dimensions: Fig.~\ref{fig:sybil} varies how many
identities the attacker registers, while Table~\ref{tab:adaptive} varies
each identity's sending rate and temporal pattern. All tested configurations
satisfy the implemented admission bounds, and authenticated p99 remains
within $72$--$73$\,ms under every defense-aware strategy. Once the
matched-volume threshold is crossed, neither malicious cloud share nor
aggregate-benign p99 grows with identity count or with the chosen pattern,
because the per-source buckets, protected capacity, priority scheduling, and
the global bucket bound admitted untrusted work independently of whether
traffic is malicious and of how that work is distributed across identities
or time.
The points in the shaded region of Fig.~\ref{fig:sybil}(a) have not
reached matched volume: their per-source buckets bind earlier, and their
higher aggregate-benign p99 reflects a different admission and rejection
mix, so they serve as lower-volume references. Containment is also costly:
unauthenticated benign rejection remains between $20\%$ and $52\%$ across
these configurations, so together the
figure and table establish bounded isolation, not immunity for every
benign client. At the nominal operating point this cost makes BE$+$G
unsuitable as-is for services with substantial anonymous or unauthenticated
demand; it fits deployments in which unauthenticated traffic can be capped or
clients can be authenticated.

\begin{table}[t]
\centering
\caption{Defense-aware traffic attacks on BE$+$G: \aaSeeds{} seeds,
measured-service queue, $K=64$. The Sybil identity-count sweep is reported
in Fig.~\ref{fig:sybil}. Rate shaping is relative to $R_u$; duty period is
$5$\,s. Duty $d$ means the attack is active for $5d$\,s of each period.
Intermediate rate-shaping and duty settings fall between these extremes
and are omitted since all tested configurations satisfy the same
admission bounds.}
\label{tab:adaptive}
\footnotesize
\setlength{\tabcolsep}{3pt}
\begin{tabular}{@{}l cccc@{}}
\toprule
\textbf{Attacker configuration} & Auth.\ p99 & Agg.\ p99 & Unauth.\ rej. & Mal.\ share \\
& (ms) & (ms) & & \\
\midrule
$0.80R_u$ ($n_{\mathrm{id}}{=}50$) & 72 & 113 & 25.1\% & 51\% \\
$1.20R_u$ ($n_{\mathrm{id}}{=}25$) & 72 & 101 & 34.6\% & 54\% \\
\addlinespace
duty $0.10$ & 73 & 122 & 20.0\% & 50\% \\
duty $1.00$ & 73 & 100 & 36.3\% & 55\% \\
\bottomrule
\end{tabular}
\end{table}

\textbf{Deadline and operating-point sensitivity.}
\label{sec:deadline}
A single p99 value at one workload and one deadline does not fully
characterize availability. This subsection first reports deadline-miss
probabilities across several latency requirements and then varies offered
request rate and queue capacity to show how the attack depends on the
system's starting operating point. The qualitative queueing effect persists
across these sweeps, while the headline amplification factor remains a
property of its operating point.

\textbf{Deadline sensitivity.}
We report $P(L>D)$ for $D\in\{50,100,200,500\}$\,ms rather than treating
$100$\,ms as a validated application requirement. Fig.~\ref{fig:deadline}
uses a Sybil flood at $\varphi=0.5$ and pools samples over $12$ seeds.
Its first three panels concern admitted cloud requests, as defined in
\S\ref{sec:eval}. %

Without defense, admitted aggregate-benign requests miss $100$\,ms
\dmAggNoneHundred{} of the time (\dmAggNoneZeroHundred{} without attack),
and $500$\,ms \dmAggNoneFiveHundred{} of the time.
BE lowers these to \dmAggBEHundred{} and \dmAggBEFiveHundred{};
BE$+$G gives \dmAggGtbHundred{} and \dmAggGtbFiveHundred{}.
These improvements must be read alongside the rejected offloads in
Table~\ref{tab:defenses}.

The ``all benign'' panel reconstructs all benign requests, including local
and rejected-offload fallback answers at the measured $1.30$\,ms edge latency.
For a deadline $D$ above that edge latency and benign offload rejection
fraction $d$, its miss probability is
$P_{\mathrm{all\,benign}}(L>D)=q_b(1-d)P_{\mathrm{admitted\,benign}}(L>D)$.
This uses benign $q_b$, not the mixture of benign and adversarial offload
probabilities.

Expected benign accuracy is likewise reconstructed from the route fractions:
local answers use clean accepted-edge correctness, admitted offloads use clean
cloud correctness conditional on offloading, and rejected offloads use
clean edge correctness conditional on offloading.
The resulting estimates are \dmAllAccNone{} without defense,
\dmAllAccBE{} with BE, and \dmAllAccGtb{} with BE$+$G.
These are model-derived expectations under fixed conditional-correctness
assumptions, not hardware-measured accuracies; lower latency from fallback
does not imply preserved prediction quality.

\textbf{Operating-point sensitivity.}
We vary offered application rate over $30$, $40$, $50$, and $60$\,req/s
and capacity over $K\in\{32,64,128\}$, each with \revCPUSeeds{} paired
clean/attack seeds, \revCPUHorizon{}\,s with \revCPUWarmup{}\,s warm-up,
the same measured service distribution, no defense, and concentrated
traffic at $\varphi\in\{0,0.5\}$ with the saturated $q_a=1$; these are
additional runs alongside the main $3000$\,s experiment, reported as mean
paired per-seed p99 ratios with pointwise $95\%$ $t$ intervals.

At $30$ and $40$\,req/s the ratios are $1.60$ and $2.66$.
At $50$\,req/s, varying $K$ changes the ratio from $5.16$ to $20.30$.
At $60$\,req/s, approximately $16.4$--$16.6\%$ of benign offloads are
rejected, so the admitted-request tail alone understates service loss.
The \benignPninetynineAmp{} headline is therefore an operating-point-specific
illustration. These sweeps still use one serial worker; they do not validate
batching or autoscaling.

\subsection{Trade-Offs and Limitations}
\label{sec:policy-tradeoff}
The global untrusted bucket is the principal mechanism that converts
per-identity limits into an identity-independent bound, but its refill rate
also determines how much legitimate unauthenticated traffic is rejected. We
sweep this rate and compare a global-bucket-only FIFO policy with the
complete BE$+$G composition to separate aggregate containment from the
additional value of protected capacity and trusted priority; the resulting
configurations map the empirical latency--rejection--accuracy trade-off
across evaluated policy choices. The sweep uses refill rates $8$, $12$,
$14.95$, $18$, $22$, $26$, and $32$/s and burst depth $20$. Trusted clients bypass this global bucket in both policies.
The bucket-only baseline has no per-source budget, protected capacity, or
priority. We retain no-defense, strict-priority, and BE references and all
tested rates (Table~\ref{tab:revision-policies}).

\begin{table}[!t]
\centering
\caption{Additional policy sweep (representative rows only; the complete
seven-rate sweep is visualized in Figs.~\ref{fig:be-mechanisms} and
\ref{fig:revision-tradeoff}), using the same cloud-work-matched Sybil
construction as Table~\ref{tab:defenses} (\S\ref{sec:eval}): eight seeds,
$800$\,s, $K=64$. T/U/A are authenticated, unauthenticated, and aggregate
benign. p99 excludes rejected offloads; rejection uses each class's
offload attempts; malicious share uses cloud service starts. The accuracy
column is the modeled all-benign fallback accuracy of \S\ref{sec:defense}
(not an end-to-end measurement). G applies only to untrusted
clients. The three reference baselines have no global bucket; the
remaining rows show the nominal ($14.95$/s) and the two extreme tested
refill rates for ``G only'' vs.\ ``BE$+$G''.}
\label{tab:revision-policies}
\footnotesize
\setlength{\tabcolsep}{2pt}
\begin{tabular}{lrrrrrrrc}
\toprule
Policy & $R_g$/s & \multicolumn{3}{c}{p99 (ms)} & \multicolumn{2}{c}{Rej.(\%)} & Mal.(\%) & Mod.\ acc.(\%) \\
 & & T & U & A & T & U & & \\
\midrule
None (FIFO) & -- & 1735.7 & 1736.5 & 1736.4 & 0.87 & 0.88 & 68.54 & 89.69 \\
Strict priority & -- & 76.9 & 6165.7 & 5279.1 & 0.87 & 0.88 & 68.53 & 89.69 \\
BE & -- & 76.6 & 1735.6 & 1543.5 & 1.19 & 2.75 & 68.68 & 89.51 \\
G (FIFO) & 8 & 77.2 & 63.0 & 75.4 & 0.00 & 74.23 & 47.82 & 83.81 \\
G (FIFO) & 14.95 & 96.7 & 86.4 & 94.3 & 0.00 & 51.86 & 59.12 & 85.63 \\
G (FIFO) & 32 & 1623.2 & 1611.8 & 1618.6 & 0.17 & 1.30 & 68.39 & 89.71 \\
BE$+$G & 8 & 68.8 & 75.9 & 70.2 & 1.35 & 74.41 & 48.09 & 83.70 \\
BE$+$G & 14.95 & 72.3 & 110.0 & 87.3 & 1.25 & 51.74 & 59.13 & 85.54 \\
BE$+$G & 32 & 76.2 & 1755.2 & 1548.7 & 1.16 & 3.07 & 68.45 & 89.49 \\
\bottomrule
\end{tabular}
\end{table}

\begin{figure*}[!t]
\centering
\includegraphics[width=0.8\textwidth]{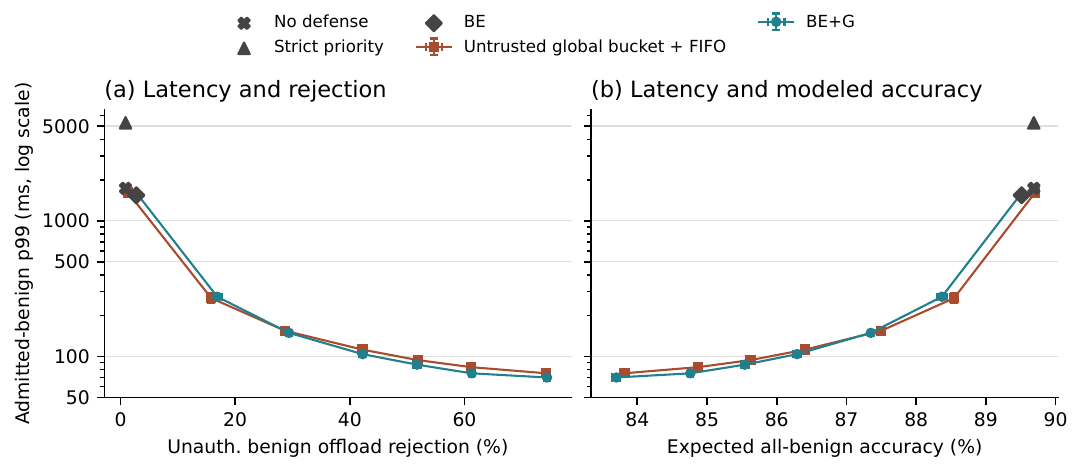}
\caption{Admission trade-offs for all tested global-bucket rates; error bars
are pointwise $95\%$ $t$ intervals across eight seeds.
Lower admitted latency is accompanied by more rejected benign offloads and
lower modeled all-benign fallback accuracy.}
\label{fig:revision-tradeoff}
\end{figure*}

Table~\ref{tab:revision-policies}
reports, for each of the $9$ representative rows (three baselines without
a global bucket, plus the nominal rate and two extremes crossed with
G-only versus BE$+$G),
separate p99 values for authenticated (T), unauthenticated (U), and
aggregate (A) benign traffic alongside rejection and modeled accuracy.
Fig.~\ref{fig:revision-tradeoff} visualizes only the aggregate (A) column
against rejection and accuracy; it does not show the T column separately.
We flag this because the T column reveals a finding the aggregate view
cannot: authenticated p99 stays within $68$--$77$\,ms across every tested
global-bucket rate under BE$+$G, essentially decoupled from the
rejection--latency trade-off that governs the aggregate curve. The runs
use the same eight-seed, $800$\,s protocol, $K=64$, and cloud-work-matched
Sybil construction described in \S\ref{sec:defense}. (Note that it is different from the primary twelve-seed,
$3000$\,s protocol, so baseline rows are not bit-identical to Table~\ref{tab:defenses}'s Sybil column.)

\begin{figure*}[!t]
  \centering
  \includegraphics[width=0.95\textwidth]{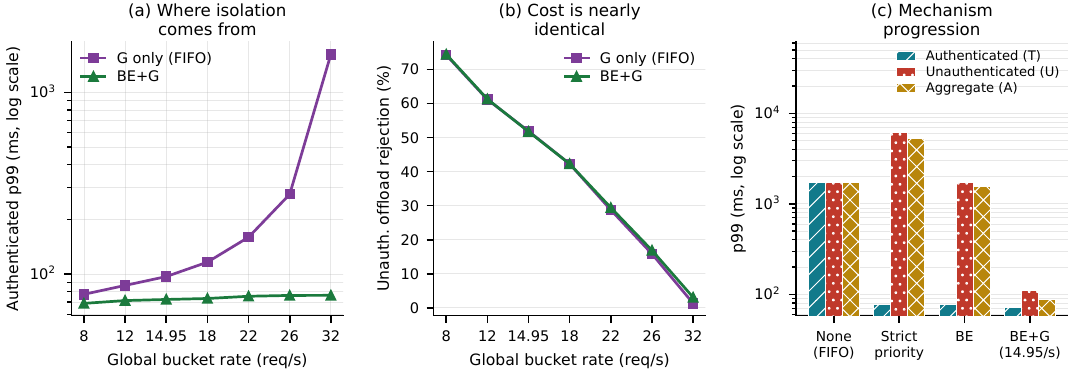}
  \caption{Global-bucket rate sweep (eight seeds, $800$\,s, $K=64$,
  cloud-work-matched Sybil construction; data from
  Table~\ref{tab:revision-policies}). (a) Authenticated p99 for G-only
  FIFO versus BE$+$G. (b) Unauthenticated offload rejection under the
  same sweep. (c) p99 by benign class across the three no-global-bucket
  baselines and the nominal BE$+$G configuration ($14.95$/s).}
  \label{fig:be-mechanisms}
\end{figure*}

Fig.~\ref{fig:be-mechanisms}(a) shows why authenticated isolation appears
only when priority scheduling and protected capacity are present. G-only
FIFO has no priority mechanism, so authenticated latency tracks the same
saturation curve as unauthenticated traffic, rising from $77.2$ to
$1623.2$\,ms as the rate relaxes from $8$ to $32$/s; BE$+$G instead keeps
authenticated p99 within $68.8$--$76.2$\,ms across the same range. This
isolation is therefore a property of the priority scheduling and
protected capacity layered on top of the global bucket, not of the global
bucket itself.

Fig.~\ref{fig:be-mechanisms}(b) shows this isolation is nearly free:
unauthenticated rejection is almost identical between the two policies
across the swept range (e.g., $51.86\%$ versus $51.74\%$ at $14.95$/s,
$15.83\%$ versus $16.86\%$ at $26$/s), diverging only at the loosest
tested rate ($32$/s: $1.30\%$ versus $3.07\%$). The complete policy is
therefore not uniformly superior, though the gap is small.

Fig.~\ref{fig:be-mechanisms}(c) traces how each mechanism contributes.
Adding priority alone (None$\to$Strict priority) drives authenticated p99
down from $1735.7$ to $76.9$\,ms, but pushes unauthenticated p99 to
$6165.7$\,ms---worse than no defense, since all queueing delay is
redirected onto lower-priority traffic. Adding per-source budgets and
protected capacity without a global bucket (Strict
priority$\to$BE) leaves authenticated p99 essentially unchanged
($76.6$\,ms) but does not repair unauthenticated p99 ($1735.6$\,ms),
since these mechanisms alone do not bound a Sybil flood spread across
many identities. Only the global bucket (BE$\to$BE$+$G at $14.95$/s)
improves both classes together: authenticated p99 remains low
($72.3$\,ms) while unauthenticated p99 falls to $110.0$\,ms.

Fig.~\ref{fig:revision-tradeoff} makes this trade-off explicit. In panel (a), each
curve sweeps the global-bucket refill rate from $8$ to $32$/s: tightening the
rate lowers admitted latency but rejects more legitimate unauthenticated
offloads, while relaxing it does the reverse. At the nominal $14.95$/s rate,
BE$+$G rejects $51.74\%$ of legitimate unauthenticated offloads for
$87.3$\,ms aggregate-benign p99; at $32$/s, rejection falls to $3.07\%$ but
p99 rises to $1548.7$\,ms.

Panel~(b) presents the same policy settings against modeled all-benign
fallback accuracy. For benign class $c\in\{\mathrm{T},\mathrm{U}\}$, let $q_c$ be its
clean offload probability, $r_c$ its rejection probability conditional on an
offload attempt, and let $a_c^{\mathrm{local}}$,
$a_c^{\mathrm{cloud}}$, and $a_c^{\mathrm{fallback}}$ denote measured clean
correctness for locally answered, cloud-offloaded, and rejected-offload
fallback predictions (i.e., conditional correctness of the
precomputed edge prediction returned after a rejected offload). 
The plotted accuracy is

{\small
\[
A_{\mathrm{benign}}
= \sum_{c\in\{T,U\}} w_c \bigl[
(1-q_c)a_c^{\mathrm{local}}
+ q_c(1-r_c)a_c^{\mathrm{cloud}}
+ q_c r_c a_c^{\mathrm{fallback}}
\bigr].
\]
}

For each benign class \(c\in\{T,U\}\), the expression partitions incoming requests into three mutually exclusive routing outcomes: local answers, admitted cloud offloads, and rejected offloads that return the previously computed edge prediction. At the matched \(\varphi=0.5\) operating point, the workload contains 25 benign principals, comprising 12 authenticated and 13 unauthenticated principals. Because every principal offers the same application-request rate, their fixed offered-benign population weights are \(w_{\mathrm{T}}=12/25\) and \(w_{\mathrm{U}}=13/25\). These weights are held constant across policies rather than recomputed from admitted requests, preventing a policy from appearing more accurate merely because it rejects more unauthenticated offloads. The nominal
BE$+$G setting reduces modeled all-benign fallback accuracy from $89.69\%$ to
$85.54\%$ while cutting aggregate-benign p99 from $1736.4$ to $87.3$\,ms.

The reference points isolate which mechanism drives this trade-off: BE
alone stays close to no defense, since per-source limits do not bound a
Sybil attack; strict priority gives the worst aggregate p99 by shifting
delay onto unauthenticated traffic; and the global-only and BE$+$G curves
nearly coincide, showing the global cap supplies most of the
aggregate-tail reduction. BE$+$G's remaining advantage---authenticated
isolation as the cap is relaxed---is visible only in
Table~\ref{tab:revision-policies}, not on this figure's aggregate-latency
axis.

\section{Conclusion}
\label{sec:conclusion}
Bounded perturbations increase offloading without increasing request volume,
producing \benignPninetynineAmp{} admitted-benign p99 amplification under
the Poisson-arrival baseline of our $50$\,req/s, serial batch-one, $K=64$
case study. Natural-input selection and
a simple global limiter provide strong attack and defense baselines,
respectively. BE$+$G preserves authenticated isolation as the global cap is
relaxed, but at the nominal operating point it rejects $52.1\%$ of legitimate
unauthenticated offloads and reduces modeled all-benign fallback accuracy from
\benignAccNoDef{} to \benignAccGtbSybil{}. Its contribution is therefore a
selectable isolation--rejection trade-off, rather than a universally superior
defense.

Although our experiments focus on EdgeBoost and one model pair, the underlying
security concern is broader: in any edge--cloud system where an
input-dependent confidence gate admits work to a scarce shared backend, that
gate also forms an admission-control boundary and should be protected by
per-identity and tier-wide limits.

\bibliographystyle{IEEEtran}
\bibliography{references}

\appendix

\subsection{Bounded Escalation Burst-Depth Provenance}
\label{app:provenance}
The per-source and global burst depths ($B_u=5$, $b_{\mathrm{gtb}}=20$) were
checked on a disjoint clean-traffic validation trace ($\varphi=0$, seeds
$100$--$111$, outside the reporting range), separately from the two
analytically-fixed refill rates. $B_u=5$ is the smallest per-source depth
keeping authenticated false rejection below $1.5\%$ (depth $4$ gives
$2.8\%$), and the reported clean-load rejection of \cleanAuthLossBE{}
is the outcome of that criterion. The global
depth has no such trade-off: any $b_{\mathrm{gtb}}\!\ge\!5$ already leaves
clean-traffic rejection at the per-source bucket's residual
${\approx}1.2\%$, so $b_{\mathrm{gtb}}=20$ is a conservative burst
allowance for flood onset rather than a value tuned against any attacked
trace.

\subsection{Clopper--Pearson Rationale}
\label{app:clopper}
Unlike the common normal (Wald) approximation, the Clopper--Pearson
interval inverts the binomial cumulative distribution directly rather than
approximating it, guaranteeing at least $95\%$ coverage of the true success
probability at the cost of being more conservative; this matters here
because several reported success rates lie near the $100\%$ boundary,
where normal approximations are known to be unreliable.

\subsection{Concentrated and Sybil Flood Arithmetic}
\label{app:flood}
Of the $50$ total principals ($25$ benign, $25$ adversarial), a
concentrated flood has each of the $25$ adversarial principals present
itself as a single identity offloading at rate $1.0$/s, so adversarial
identities are $25$ of $50$ visible identities ($50\%$) and jointly offer
$25{\times}1.0{=}25$ adversarial offloads/s. A Sybil flood instead has each
of the same $25$ adversarial principals split into $10$ separate
identities (matching the $n_{\mathrm{id}}=250$ case used in
\S\ref{sec:adaptive}), each offloading at only $0.1$/s; adversarial
identities are now $250$ of $275$ visible identities ($90.9\%$), while the
aggregate adversarial offload rate is unchanged at $250{\times}0.1{=}25$/s.

\subsection{Protocol Differences Across Repeated-Looking Results}
\label{app:protocols}
Several results that appear to repeat one another are in fact produced by
distinct protocols and should not be treated as independent replications
of a single population.

The \benignPninetynineAmp{} headline is reported by three separately
configured runs whose tail ratios differ: the primary $3000$\,s, $12$-seed
run gives \cleanBenignPninetynine{}$\to$\attackBenignPninetynine{}
(\benignPninetynineAmp{}); the fixed-volume comparison,
configured with its own attacker-designation
construction, gives \cfPnnClean{}$\to$\cfPnnPert{}\,ms ($10.53\times$);
and the shorter \revCPUHorizon{}\,s,
\revCPUSeeds{}-seed sensitivity run at $\Lambda=50$, $K=64$ gives a paired
per-seed ratio of $10.28$ $[9.88,10.68]$ (operating-point sensitivity,
\S\ref{sec:tradeoff}). These
runs share the simulator and service distribution and differ in horizon,
seed count, traffic construction, and whether ratios are formed per seed
or from pooled summaries.

The multi-gate test (Table~\ref{tab:cascade}) uses $1000$ images with
separately frozen thresholds; the budget sweep
(Fig.~\ref{fig:epssweep}) and the preprocessing evaluation
(Fig.~\ref{fig:preproc}) use the full $1000$ test images with three
attack seeds and one primary plus two robustness seeds, respectively.
These protocols yield different operating points and should not be
compared as repeated measurements of one population.

\end{document}